\documentclass[journal]{IEEEtran}

\usepackage{xcolor,soul,framed} 
\colorlet{shadecolor}{yellow}
\usepackage{graphicx}
\graphicspath{{../pdf/}{../jpeg/}}
\DeclareGraphicsExtensions{.pdf,.jpeg,.png}
\usepackage[cmex10]{amsmath}
\usepackage{latexsym,bm,amssymb}
\usepackage{array}
\usepackage{mdwmath}
\usepackage{mdwtab}
\usepackage{eqparbox}
\usepackage{url}
\usepackage{enumerate}
\usepackage{amsfonts}
\usepackage{algorithmic}
\usepackage{algorithm}
\usepackage[numbers,sort&compress]{natbib}
\usepackage[justification=raggedright,singlelinecheck=false]{caption} 
\usepackage{mathrsfs}
\usepackage{diagbox}
\usepackage{graphicx}
\usepackage{tabularx}
\usepackage{subcaption}
\usepackage{caption}
\usepackage[justification=raggedright,singlelinecheck=false]{caption} 
\usepackage{tabularx} 

\usepackage{multirow}

\usepackage{booktabs} 

\usepackage{footnote}

\usepackage{nomencl}
\makenomenclature
\usepackage{ifthen}
  \renewcommand{\nomgroup}[1]{%
  \item[\bfseries
  \ifthenelse{\equal{#1}{A}}{Symbols of UC}{%
  \ifthenelse{\equal{#1}{S}}{Symbols of CL}{}}%
  ]}

\renewcommand{\baselinestretch}{1}

\usepackage{float}
\begin{document}

\captionsetup{font={small}}

\title{Impact and Mitigation of Current Saturation Algorithms in Grid-Forming Inverters on Power Swing Detection}

\author{Yanshu Niu,~\IEEEmembership{Student Member,~IEEE,} Zhe Yang,~\IEEEmembership{Member,~IEEE,} and Bikash C. Pal,~\IEEEmembership{Fellow,~IEEE}

}

\markboth{}%
{Shell \MakeLowercase{\textit{et al.}}: Bare Demo of IEEEtran.cls for IEEE Journals}
{}

\maketitle

\begin{abstract}
Grid-forming (GFM) inverter-based resources (IBRs) are capable of emulating the external characteristics of synchronous generators (SGs) through the careful design of the control loops. However, the current limiter in the control loops of the GFM IBR poses challenges to the effectiveness of power swing detection functions designed for SG-based systems. Among various current limiting strategies, current saturation algorithms (CSAs) are widely employed for their strict current limiting capability, and are the focus of this paper. The paper presents a theoretical analysis of the conditions for entering and exiting the current saturation mode of the GFM IBR under three CSAs. The corresponding impedance trajectories observed by the relay on the GFM IBR side are investigated. The analysis results reveal that the unique impedance trajectories under these CSAs markedly differ from those associated with SGs. Moreover, it is demonstrated that the conventional power swing detection scheme may lose functionality due to the rapid movement of the trajectory. To mitigate this issue, an optimal current saturation strategy is proposed. Conclusions are validated through simulations in MATLAB/Simulink.
\end{abstract}

\begin{IEEEkeywords}
grid-forming control, current limiter, current saturation algorithm, power swing detection.
\end{IEEEkeywords}

\printnomenclature



\IEEEpeerreviewmaketitle

\section{Introduction}\label{introduction}
\IEEEPARstart{T}{o} achieve the net-zero commitment, inverter-based resources (IBRs) driven by renewable energy, such as wind and solar, are gradually replacing synchronous generators (SGs) powered by fossil fuels in power systems~\cite{iea2024}. Although grid-following (GFL) remains the predominant control method for IBRs, grid-forming (GFM) more accurately mimics the rotor dynamic response of synchronous generators (SGs)~\cite{lasseter2020}, on which conventional protection schemes are based. Therefore, compared to GFL IBRs, GFM IBRs are considered more effective in alleviating the potential risk of malfunctions in power swing detection functions as the penetration of IBRs increases~\cite{lin2020research}. However, differing from SGs, the external response of GFM IBRs is governed by control loops~\cite{yang2024protection}, lacking a rigid rotor body to provide inherent inertia. Moreover, the current limiter in the control loop restricts the current reference to protect the power electronic components. Consequently, how GFM IBRs influence power swing detection functions remains an unsolved mystery~\cite{lin2020research,munz2024}, and further research is required to investigate the underlying mechanisms.

Power swing detection serves two primary functions: power swing blocking (PSB) and out-of-step tripping (OST). The PSB function aims to discriminate between a fault and a power swing, blocking the distance protection to prevent unnecessary tripping in a power swing event~\cite{Hou2005}. The objective of the OST function is to distinguish between stable and unstable power swings, separating the system at pre-determined network locations to prevent further cascading faults in an unstable power swing event~\cite{psrc2005,nerc2013}.
Mature power swing detection methods include traditional approaches based on the rate of change of impedance or resistance, as well as non-traditional techniques developed for microprocessor-based relays~\cite{psrc2005,nerc2013,haddadi2017,fischer2012tutorial}. Among them, those utilising the rate of change of apparent impedance remain the most widely implemented in commercial relays, with blinder-based schemes being the predominant realisation~\cite{ge2023,rel670manual,siemens2023,ieee2016}. In practice, three sets of blinders, including inner, middle, and outer, are commonly applied~\cite{rel670manual,Haddadi2019}, enabling the simultaneous implementation of both PSB and OST functions. However, the effectiveness of this method for power swing detection in power systems dominated by GFM IBRs requires further investigation.

To prevent overcurrent from causing damage to the power electronic components in GFM IBRs, current saturation algorithms (CSAs) and virtual impedance (VI) methods are widely applied to limit the current~\cite{fan2022, qoria2020, Rokrok2022}. This paper focuses on three typical CSAs: circular CSA, d-axis priority CSA, and q-axis priority CSA. Limited research has investigated the conditions for exiting the current saturation mode. The authors in~\cite{qoria2020} assume that the current exits the saturation mode at the intersection point of the \(P-\delta\) curves under saturated and unsaturated modes. Although the simulation-based case studies in~\cite{qoria2020} seem to verify this assumption, a theoretical analysis is lacking to prove it. 
The fault recovery process under the constant CSA has been analysed in~\cite{fan2023}, revealing different conditions for transitions between the normal operating mode and the current saturation mode. However, further theoretical analysis and a closed-form expression are not provided. It is highlighted in~\cite{li2023,lu2024,arjomandi2024} that IBRs can potentially fall into saturated stable equilibrium points (SSEPs) during fault recovery, and the conditions for exiting the current saturation mode are also derived. However, the analysis only addresses the user-defined current angle in constant CSA, without considering the other three types of CSAs. The exiting conditions under other different CSAs, which remain not well understood, may significantly impact the apparent impedance trajectories.

Some research has preliminarily explored the impact of replacing SGs with renewable energy resources or IBRs on power swing detection. The influence of Type-III wind turbine generators on power swing detection was investigated in~\cite{Haddadi2019}, and the scope was further extended to more general IBRs in~\cite{Haddadi2021}. However, these analyses are based solely on simulations, without providing theoretical explanations. A method is proposed to discriminate between symmetrical faults and power swings in~\cite{Rao2022}, but the research scope remains limited to Type-III wind turbine generators. Simulations presented in~\cite{Jayamohan2023} demonstrate that the rate of change of apparent impedance during stable and unstable power swings is significantly faster in GFL IBR-based systems than in systems with only SGs. This rapid rate of change poses a risk of PSB and OST malfunctions. From a more theoretical perspective, a dynamic model is developed in~\cite{nasr2024part1} to illustrate that the DC-link voltage control dynamics can significantly amplify the rate of impedance change~\cite{nasr2024part2}. The authors of~\cite{xiong2023} and~\cite{xiong2024} conducted a detailed theoretical analysis of power swing trajectories caused by IBRs and qualitatively investigated the influence of control parameters. However, the analyses in~\cite{Jayamohan2023,nasr2024part1,nasr2024part2,xiong2023,xiong2024} are limited to GFL IBRs and do not address GFM IBRs. Although GFM IBRs are included in the research scope in~\cite{Xiong2023unlimited}, it does not account for the role of the current limiter. To the best of the authors' knowledge, no previous study has theoretically analysed the power swing trajectories of GFM IBRs under CSA-based current-limiting strategies.

To fill this gap, the apparent impedance trajectories of the GFM IBR during power swings under three types of CSA are analysed in this paper. Furthermore, the impact of these trajectories on legacy power swing detection protection is investigated. The main contributions of this paper are:
\begin{itemize}
\item The conditions for entering and exiting current saturation mode are derived for three types of CSAs: circular, d-axis priority, and q-axis priority. The analysis reveals that the entering and exiting angle sets are not complementary to each other.
\item The full-cycle power swing trajectories under these three CSA strategies are examined. The analysis highlights their distinct characteristics compared to those of conventional SG-based systems.
\item An optimal current saturation strategy is proposed to improve the reliability of legacy power swing detection by ensuring that the impedance trajectory moves continuously during the transition from the unsaturated to the saturated mode.
\end{itemize}
The theoretical analysis is validated through simulation models implemented on the MATLAB/Simulink platform.

The paper is organised as follows. In Section~\ref{System Model}, the system model is elaborated. In Section~\ref{Current Saturation Algorithm}, three typical CSAs are introduced, and the conditions for entering and exiting current saturation mode are derived. Section~\ref{Analysis of Power Swing Trajectories} discusses the power swing trajectories with the CSAs. The optimal current saturation strategy is proposed in Section~\ref{Optimal Current Saturation Strategy}. Section~\ref{Simulations and Case Studies} presents the simulation and case studies to validate the theoretical analysis. Finally, Section~\ref{Conclusion} concludes the paper.
\section{System Model}
\label{System Model}
\begin{figure*}[htbp]
    \centering
    \begin{subfigure}{0.9\textwidth}
        \centering
        \includegraphics[width=\textwidth, clip]{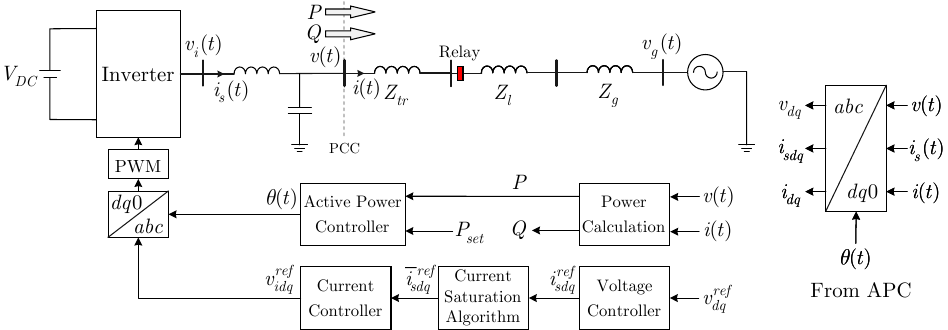}
        \captionsetup{justification=centering} 
        \caption{} 
        \label{fig:System_Model_System_Model}
    \end{subfigure}
    \hfill
    \begin{subfigure}{0.9\textwidth}
        \centering
        \includegraphics[width=\textwidth, clip]{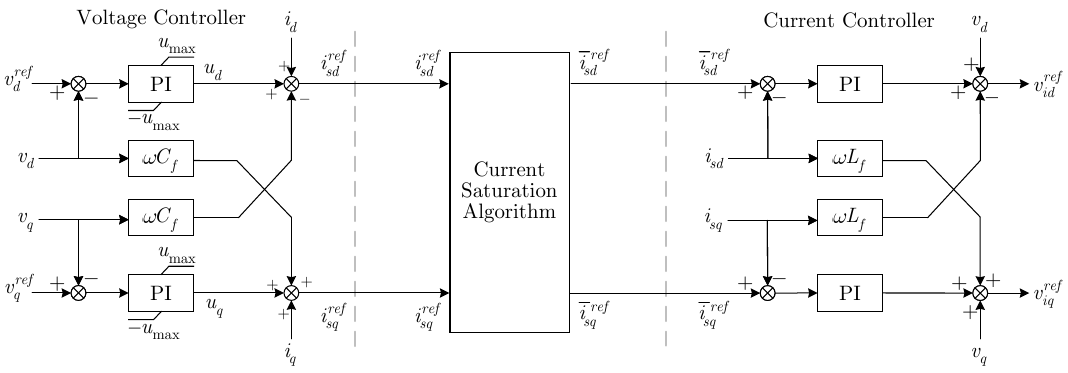}
        \captionsetup{justification=centering} 
        \caption{} 
        \label{fig:System_Model_Control_Loops}
    \end{subfigure}
    \captionsetup{justification=raggedright,singlelinecheck=false} 
    \caption{Grid-connected GFM IBR system. (a) System model and control structure. (b) Control block diagram.}
    \label{fig:System_Model}
\end{figure*}

Fig.~\ref{fig:System_Model}(a) shows the system configuration of a grid-connected GFM IBR system. The DC source is connected to the grid through an inverter, an LC filter, a step-up transformer \(Z_{tr}\) and a power line \(Z_{l}\). The grid is modeled as a Thevenin equivalent voltage source \({V}_{g}\angle\theta_{g}\) in series with an impedance \(Z_{g}\). The low-voltage side of the transformer is considered as the point of common coupling (PCC). Thus, the total impedance between the PCC and the Thevenin equivalent voltage source of the grid is \(Z_{tr}+Z_{l}+Z_{g}=Z_{T}\angle\phi=R_{T}+jX_{T}\). The voltage phasor angle of the GFM IBR, \(\theta\), is generated by the active power controller (APC) in the virtual synchronous machine control scheme to emulate the swing equations of the SGs; while the voltage set point of the GFM IBR should be determined by the reactive power controller (RPC). However, as RPC is not the focus of this paper, the voltage set point is assumed to be \( \dot{V}^{\text{ref}} = v_{d}^{\text{ref}} + jv_{q}^{\text{ref}} = 1 + j0 \). With the local voltage reference of the GFM IBR aligned along the d-axis, where \(\theta=0\), the power angle between the IBR and the grid is defined by \(\delta=\theta-\theta_{g}\), i.e., \(\delta=-\theta_{g}\).  A distance protection relay is installed at the power line terminal near the transformer, equipped with both three-zone mho-type distance protection and power swing detection functions.

\begin{figure}[htbp]
    \centering
    \begin{subfigure}{0.47\columnwidth}
        \centering
        \includegraphics[width=\textwidth]{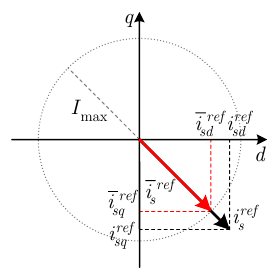}
        \captionsetup{justification=centering} 
        \caption{} 
        \label{fig:Circular_Priority}
    \end{subfigure}
    \hfill
    \begin{subfigure}{0.47\columnwidth}
        \centering
        \includegraphics[width=\textwidth]{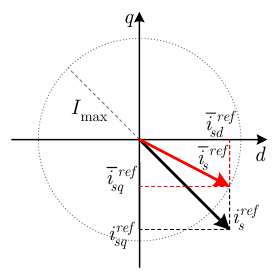}
        \captionsetup{justification=centering} 
        \caption{} 
        \label{fig:D_Axis_Priority}
    \end{subfigure}
    \hfill
    \begin{subfigure}{0.47\columnwidth}
        \centering
        \includegraphics[width=\textwidth]{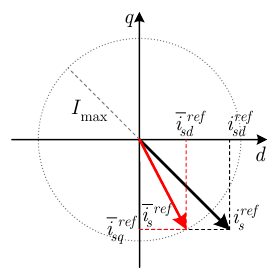}
        \captionsetup{justification=centering} 
        \caption{} 
        \label{fig:Q_Axis_Priority}
    \end{subfigure}

    \captionsetup{justification=raggedright,singlelinecheck=false} 
    \caption{The new current references generated by the CSAs. (a) Circular. (b) D-Axis Priority. (c) Q-Axis Priority.}
    \label{fig:Three_Types_CSAs}
\end{figure}

A typical voltage-current dual-loop cascaded control structure is implemented in this paper, as illustrated in Fig.~\ref{fig:System_Model}(b). To prevent overcurrent in any abnormal operating condition from damaging the power electronic components in the inverter, a CSA is essential. It limits the magnitude of the current reference output from the voltage controller to the maximum allowable value \(I_{\text{max}}\), generating a new current reference as the input to the current controller. To prevent integrator windup, the clamping anti-windup technique is employed within the voltage PI controllers to limit the magnitude of the PI output to \(u_{\text{max}}\). The CSA can be implemented by prioritising the current vector angle, d-axis current and q-axis current, as presented in Fig.~\ref{fig:Three_Types_CSAs}. These different algorithms have distinct conditions for exiting current saturation, which are reflected in different apparent impedance trajectories and will be investigated in the next sections.
\section{Current Saturation Algorithm}
\label{Current Saturation Algorithm}
In this section, the condition for entering current saturation is derived, followed by the analysis of the conditions for exiting saturation under different CSAs.
\vspace{-0.5em}
\subsection{Condition for Entering Current Saturation}
Independent of the specific CSA employed, the condition for entering current saturation remains consistent and can be expressed as:
\begin{equation}
\label{eq:Enter_Saturation_Criterion}
i_{sd}^2+i_{sq}^2 >I_{\text{max}}^2 .
\end{equation}
Based on~(\ref{eq:Enter_Saturation_Criterion}) and Kirchhoff's voltage law, the GFM IBR enters current saturation~\cite{wang2023transient}, if
\begin{equation}
\label{eq:Enter Saturation Angle}
\cos\delta < \frac{ \left(v_{d}^{\text{ref}}\right)^{2}+\left(V_{g}\right)^{2}-\left(| Z_{T} | I_{\text{max}} \right)^2}{2 v_{d}^{\text{ref}} V_{g}}.
\end{equation}
The power angle at the critical condition where the current transitions from unsaturated to saturated is defined as
\begin{equation}
\label{eq:Enter Saturation Angle defination}
\delta_{\text{enter}} = \arccos \left[  \frac{ \left(v_{d}^{\text{ref}}\right)^{2}+\left(V_{g}\right)^{2}-\left(| Z_{T} | I_{\text{max}} \right)^2}{2 v_{d}^{\text{ref}} V_{g}} \right].
\end{equation}
Thus, when
\begin{equation}
\delta \in [\delta_{\text{enter}},2\pi - \delta_{\text{enter}}],
\end{equation}
the power angle satisfies the condition for current saturation.
\subsection{Conditions for Exiting Current Saturation}
The conditions for exiting current saturation are not inherently complementary to those for entering saturation. Instead, they depend on the CSAs, as will be elaborated in the following discussion.
\subsubsection{Circular Current Saturation Algorithm}
\begin{figure}[!t]
    \vspace{-0.7em}
    \centering
    \begin{subfigure}{0.47\columnwidth}
        \centering
        \includegraphics[width=\textwidth]{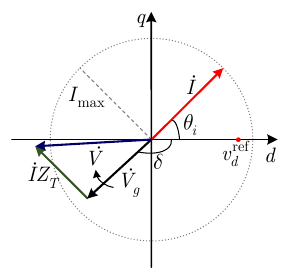}
        \captionsetup{justification=centering} 
        \caption{}
        \label{fig:Explaining_current_sign_before}
    \end{subfigure}
    \hfill
    \begin{subfigure}{0.47\columnwidth}
        \centering
        \includegraphics[width=\textwidth]{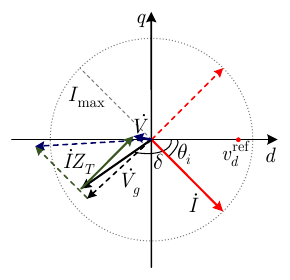}
        \captionsetup{justification=centering} 
        \caption{}
        \label{fig:Explaining_current_sign_after}
    \end{subfigure}
    \captionsetup{justification=raggedright,singlelinecheck=false} 
    \caption{The process of a change in the sign of \(\bar{i}_{sq}\). (a) Before \(v_{q}\) tracks \(v_{q}^{\text{ref}}\). (b) After \(v_{q}\) tracks \(v_{q}^{\text{ref}}\), as indicated by the solid lines.}
    \label{fig:Explaining_current_sign}
\end{figure}

Fig.~\ref{fig:Three_Types_CSAs}(a) illustrates the priority of the current angle, referred to as the circular CSA, which limits the magnitude to \(I_{\text{max}}\) while maintaining the angle consistent with the unsaturated current reference before the current limiter. Under this algorithm, the generated d-axis and q-axis current references are
\begin{align}
\bar{i}_{sd}^{\text{ref}} &= \frac{i_{sd}^{\text{ref}}}{|i_{sd}^{\text{ref}}|} \times 
\min\left( 
|i_{sd}^{\text{ref}}|, 
|i_{sd}^{\text{ref}}| \times \frac{I_\text{max}}{\sqrt{\left(i_{sd}^{\text{ref}}\right)^2 + \left(i_{sq}^{\text{ref}}\right)^2}}
\right) 
\label{eq:isd_ref_Circular}  \\ 
\bar{i}_{sq}^{\text{ref}} &= \frac{i_{sq}^{\text{ref}}}{|i_{sq}^{\text{ref}}|} \times 
\min\left( 
|i_{sq}^{\text{ref}}|, 
|i_{sq}^{\text{ref}}| \times \frac{I_\text{max}}{\sqrt{\left(i_{sd}^{\text{ref}}\right)^2 + \left(i_{sq}^{\text{ref}}\right)^2}}
\right).
\label{eq:isq_ref_Circular} 
\end{align}
When the current enters the saturation mode, \(v_{d}\) and \(v_{q}\) cannot track their references, causing the rapid clamping of the PI controllers in the voltage control loops. Thus, the current references for the d-axis and q-axis before the CSA are
\begin{align}
i_{sd}^{\text{ref}} &= \text{sgn}\left(u_{d}\right)u_{\text{max}}+\bar{i}_{sd} \label{eq:isd_ref_saturation} \\
i_{sq}^{\text{ref}} &= \text{sgn}\left(u_{q}\right)u_{\text{max}}+\bar{i}_{sq}, \label{eq:isq_ref_saturation}
\end{align}
where \(\bar{i}_{sd}=\bar{i}_{d}-v_{q}\omega C_{f} \) and \(\bar{i}_{sq}=\bar{i}_{q}+v_{d}\omega C_{f}\).
Assuming that the response of the current controller is fast enough to accurately track the references, i.e., \(\bar{i}_{sdq}=\bar{i}_{sdq}^{\text{ref}}\), it can be derived that 
\begin{align}
\left(\frac{1}{k}-1\right)\bar{i}_{sd}^{\text{ref}} &= \text{sgn}\left(u_{d}\right)u_{\text{max}} \\
\left(\frac{1}{k}-1\right)\bar{i}_{sq}^{\text{ref}} &= \text{sgn}\left(u_{q}\right)u_{\text{max}},
\end{align}
where \(k={I_\text{max}}/{\sqrt{\left(i_{sd}^{\text{ref}}\right)^2 + \left(i_{sq}^{\text{ref}}\right)^2}}\). The sign of \(u_{d}\) depends on the sign of \((v_{d}^{\text{ref}}-v_{d})\), according to the voltage control loop in Fig.~\ref{fig:System_Model}(b); similarly, the sign of \(u_{q}\) depends on the sign of \((v_{q}^{\text{ref}}-v_{q})\), where \(v_{d}\) and \(v_{q}\) refer to the d-axis and q-axis components of the PCC voltage, denote as
\begin{align}
v_{d} &= \hphantom{-}V_{g}\cos\delta + |Z_{T}|\left(\bar{i}_{sd}\cos\phi-\bar{i}_{sq}\sin\phi\right) ,\label{eq:vd} \\
v_{q} &= -V_{g}\sin\delta + |Z_{T}|\left(\bar{i}_{sd}\sin\phi+\bar{i}_{sq}\cos\phi\right). \label{eq:vq}
\end{align}
Since \(k\leq1\) under the saturation mode, the magnitude of the d-axis and q-axis currents satisfy
\begin{equation}
|\bar{i}_{sd}|=|\bar{i}_{sq}|=\frac{\sqrt{2}}{2}I_{\text{max}}.
\end{equation}
The signs of \(\bar{i}_{sd}\) and \(\bar{i}_{sq}\) are consistent with those of \(u_{d}\) and \(u_{q}\), respectively. Therefore, when the current has not exited saturation, that is, when \(v_{d}\) and \(v_{q}\) fail to simultaneously track their respective references, the signs of \(\bar{i}_{sd}\) or \(\bar{i}_{sq}\) may change with the variation of the phase angle \(\theta_{g}\) of \(V_{g}\), as illustrated in the example shown in Fig.~\ref{fig:Explaining_current_sign}.
As the power angle \(\delta\) increases in Fig.~\ref{fig:Explaining_current_sign}(a), the grid voltage \(\dot{V}_{g}\) rotates counterclockwise. Before \(v_{q}\) can track the reference \(v_{q}^{\text{ref}}\), \(v_{q}^{\text{ref}}-v_{q}\) is positive, resulting in a positive \(\bar{i}_{sq}\). Similarly, \(\bar{i}_{sd}\) remains positive. Thus, the saturated current is located in the first quadrant. As \(\dot{V}_g\) continues to rotate, moving the phasor \(\dot{V}\) into the second quadrant as shown in Fig.~\ref{fig:Explaining_current_sign}(b), \(v_q^{\text{ref}} - v_q\) changes from positive to negative, causing \(\bar{i}_{sq}\) to also become negative, while \(\bar{i}_{sd}\) remains positive. Consequently, the saturated current rapidly changes from the first quadrant to the fourth quadrant. This change in the current leads to a significant reduction in the magnitude of $\dot{V}$, as shown by the change in $\dot{V}$ from Fig.~\ref{fig:Explaining_current_sign}(a) to Fig.~\ref{fig:Explaining_current_sign}(b).

\begin{figure}[!t]
    \vspace{-0.7em}
    \centering
    \begin{subfigure}{0.478\columnwidth}
        \centering
        \includegraphics[width=\textwidth]{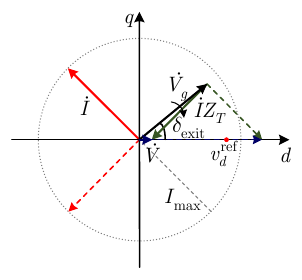}
        \captionsetup{justification=centering} 
        \caption{}
        \label{fig:Circular_Exiting_Case1}
    \end{subfigure}
    \hfill
    \begin{subfigure}{0.478\columnwidth}
        \centering
        \includegraphics[width=\textwidth]{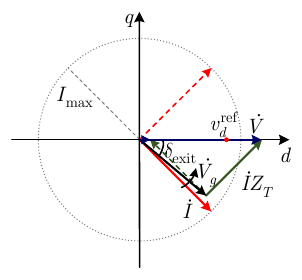}
        \captionsetup{justification=centering} 
        \caption{}
        \label{fig:Circular_Exiting_Case2}
    \end{subfigure}

    \captionsetup{justification=raggedright,singlelinecheck=false} 
    \caption{Two scenarios for current exiting saturation under circular CSA control. The dashed lines represent the moment just before \(v_{q}\) tracks \(v_{q}^{\text{ref}}\), while the solid lines represent the moment just after \(v_{q}\) tracks \(v_{q}^{\text{ref}}\). (a) \(\dot{V}_{g}\) rotates clockwise. (b) \(\dot{V}_{g}\) rotates counterclockwise.}
    \label{fig:Circular_Exiting}
\end{figure}

Since the saturated current exhibits a rapid change before and after the d-axis or q-axis voltage successfully reaches its reference, only two possible scenarios exist under circular CSA control where both d-axis and q-axis voltages simultaneously reach their relative references, as shown in Fig.~\ref{fig:Circular_Exiting}.
Just before \(v_q\) reaches \(v_q^{\text{ref}}\), the saturated current \(\dot{I}\) and the PCC voltage \(\dot{V}\) are located at the positions indicated by the red dashed line and the blue dashed line in Fig.~\ref{fig:Circular_Exiting}, respectively. Once \(v_q\) successfully reaches \(v_q^{\text{ref}}\), the sign of \(\bar{i}_{sq}\) changes, resulting \(\dot{I}\) and \(\dot{V}\) rapidly shift to the positions marked by the red solid line and the blue solid line. During this transition, \(\dot{V}\) passes through the PCC voltage set point \(v_{dq}^{\text{ref}}\), thereby satisfying the exit condition and enabling the current to exit saturation. The clockwise and counterclockwise variations of \(\delta\) shown in the figure share the same exit condition, which is
\begin{equation}
-V_{g}\sin\delta+\frac{\sqrt{2}}{2}I_{\text{max}}|Z_{T}|\cos\phi+\frac{\sqrt{2}}{2}I_{\text{max}}|Z_{T}|\sin\phi=v_{q}^{\text{ref}}=0.
\end{equation}
Defining
\begin{equation}
\delta_{\text{exit}}^{\text{circular}}=\arcsin \left[\frac{\sqrt{2}I_{\text{max}}|Z_{T}|\left(\cos\phi+\sin\phi\right)}{2V_{g}}\right],
\end{equation}
the set of power angles \(\delta\) that enable the current to exit the saturation mode under circular SCA control is given by
\begin{equation}
\delta \in [0,\delta_{\text{exit}}^{\text{circular}}] \cup[2\pi-\delta_{\text{exit}}^{\text{circular}},2\pi].
\end{equation}
\subsubsection{D-Axis Priority Current Saturation Algorithm}
With the d-axis priority current saturation algorithm shown in Fig.~\ref{fig:Three_Types_CSAs}(b), the d-axis and q-axis current references are as follows.
\begin{align}
\bar{i}_{sd}^{\text{ref}} &= \frac{i_{sd}^{\text{ref}}}{|i_{sd}^{\text{ref}}|} \times \min\left(|i_{sd}^{\text{ref}}|, I_\text{max}\right) \label{eq:isd_ref_D_Priority} \\
\bar{i}_{sq}^{\text{ref}} &= \frac{i_{sq}^{\text{ref}}}{|i_{sq}^{\text{ref}}|} \times \min\left(|i_{sq}^{\text{ref}}|, \sqrt{(I_\text{max})^2 - (\bar{i}_{sd}^{\text{ref}})^2}\right) \label{eq:isq_ref_D_Priority}.
\end{align}
After the current is limited, the GFM IBR can exit the saturation mode, if
\begin{equation}
\left(i_{sd}^{\text{ref}}\right)^2+\left(i_{sq}^{\text{ref}}\right)^2<I_{\text{max}}^2,
\end{equation}
where \(i_{sd}^{\text{ref}}\) and \(i_{sq}^{\text{ref}}\) are expressed as~(\ref{eq:isd_ref_saturation}) and~(\ref{eq:isq_ref_saturation}). With the saturation of the current, the voltages on the d-axis and q-axis can not track their respective voltage references. Consequently, the integrators within the PI controllers of the voltage loops, as shown in Fig.~\ref{fig:System_Model}(b), will clamp promptly, reaching their maximum output, \(u_{\text{max}}\). Thus, the condition for exiting the saturation has the expression as
\begin{equation}
\left [ \text{sgn}\left ( u_{d} \right ) u_{\text{max}} + \text{sgn}\left ( \bar{i}_{sd} \right ) I_{\text{max}} \right ]^2+\left [ \text{sgn}\left ( u_{q} \right ) u_{\text{max}}\right ]^2 < I_{\text{max}}^2.
\end{equation}
Assuming that \(u_{\text{max}}\ll I_{\text{max}}\), it is derived that
\begin{equation}
\text{sgn}\left ( u_{d} \right )\text{sgn}\left ( \bar{i}_{sd} \right )  < 0.
\end{equation}
There are two cases that can address the above inequality.
\begin{align}
\text{Case 1:}~& u_{d} <0,~ \bar{i}_{sd}= + I_{\text{max}};\\
\text{Case 2:}~& u_{d} >0,~ \bar{i}_{sd}= - I_{\text{max}}.
\end{align}
In Case 2, the solution does not exist; therefore, the solution in Case 1 yields a unique exit condition, which is
\begin{equation}
\cos\delta > \frac{v_{d}^{ref} - |Z_{T} |  I_{\text{max}} \cos\phi}{V_{g}}.
\end{equation}
Defining
\begin{equation}
\label{eq:Exit Saturation Angle_D-Axis}
\delta_{\text{exit}}^{\text{d-axis}} = \arccos \left[ \frac{v_{d}^{ref} - |Z_{T} |  I_{\text{max}} \cos\phi}{V_{g}} \right],
\end{equation}
the power angle set of exiting the current saturation with d-axis priority CSA is
\begin{equation}
\delta \in [0,\delta_{\text{exit}}^{\text{d-axis}}] \cup[2\pi-\delta_{\text{exit}}^{\text{d-axis}},2\pi]
\end{equation}
\subsubsection{Q-Axis Priority Current Saturation Algorithm}
If the q-axis current reference is prioritised for current limiting, as shown in Fig.~\ref{fig:Three_Types_CSAs}(c), the d-axis and q-axis references generated by the current limiter are
\begin{align}
\bar{i}_{sd}^{\text{ref}} &= \frac{i_{sd}^{\text{ref}}}{|i_{sd}^{\text{ref}}|} \times \min\left(|i_{sd}^{\text{ref}}|, \sqrt{(I_\text{max})^2 - (\bar{i}_{sq}^{\text{ref}})^2}\right) 
\label{eq:isd_ref_Q_Priority} \\
\bar{i}_{sq}^{\text{ref}} &= \frac{i_{sq}^{\text{ref}}}{|i_{sq}^{\text{ref}}|} \times \min\left(|i_{sq}^{\text{ref}}|, I_\text{max}\right) \label{eq:isq_ref_Q_Priority} 
\end{align}
The condition for current exiting saturation under q-axis priority CSA control is similar to that under d-axis priority. At the moment of exiting saturation, the signs of \(u_{q}\) and \(\bar{i}_{sq}\) satisfy the relationship
\begin{equation}
\text{sgn}\left ( u_{q} \right )\text{sgn}\left ( \bar{i}_{sq} \right )  < 0.
\end{equation}
It can be discussed in two cases:
\begin{align}
\text{Case 1:}~& u_{q} <0,~  \bar{i}_{sq} = + I_{\text{max}};\\
\text{Case 2:}~& u_{q} >0,~  \bar{i}_{sq} = - I_{\text{max}}.
\end{align}
The solutions corresponding to the two cases are
\begin{equation}
\sin\delta > -\frac{  |Z_{T} |  I_{\text{max}} \cos\phi}{V_{g}}
\end{equation}
and
\begin{equation}
\sin\delta < \frac{ |Z_{T} |  I_{\text{max}} \cos\phi}{V_{g}},
\end{equation}
respectively. Defining 
\begin{equation}
\label{eq:Exit Saturation Angle_Q-Axis}
\delta_{\text{exit}}^{\text{q-axis}} = \arcsin \left[ \frac{ |Z_{T} |  I_{\text{max}} \cos\phi}{V_{g}} \right],
\end{equation}
the current can exit the saturation mode, if the power angle belongs to the set
\begin{equation}
\delta \in [0, \delta_{\text{exit}}^{\text{q-axis}}] \cup [\pi-\delta_{\text{exit}}^{\text{q-axis}}, \pi+\delta_{\text{exit}}^{\text{q-axis}}] \cup [2\pi-\delta_{\text{exit}}^{\text{q-axis}}, 2\pi] 
\end{equation}
However, it is worth noting that the exit angle set \(\delta \in[\pi-\delta_{\text{exit}}^{\text{q-axis}}, \pi+\delta_{\text{exit}}^{\text{q-axis}}]\) also belongs to the saturation angle set. When the power angle is within this set, both the entering and the exiting saturation conditions are simultaneously satisfied, causing the IBR to oscillate between the two modes, which negatively impacts system stability. To eliminate the oscillation, within this subset, forced saturation is implemented as follows. 
\begin{align}
\bar{i}_{sq} &= +I_{\text{max}}, & \text{if } \delta \in [\pi - \delta_{\text{exit}}^{\text{q-axis}}, \pi]; \\
\bar{i}_{sq} &= -I_{\text{max}}, & \text{if } \delta \in [\pi, \pi + \delta_{\text{exit}}^{\text{q-axis}}].
\end{align}
Therefore, the set of power angles for exiting saturation under q-axis priority CSA control is
\begin{equation}
\delta \in [0, \delta_{\text{exit}}^{\text{q-axis}}] \cup [2\pi-\delta_{\text{exit}}^{\text{q-axis}}, 2\pi].
\end{equation}
\section{Analysis of Power Swing Trajectories}
\label{Analysis of Power Swing Trajectories}
\begin{figure}[htbp]
\vspace{-0.7em}
\centering
\includegraphics[width=3in]{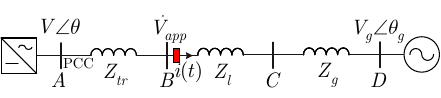}
\caption{A Single Machine GFM IBR Grid-Connected System.}
\label{fig:Single Machine Grid Connected Circuit Model}
\end{figure}

The power swing trajectories can be affected by different CSAs. In this section, the trajectory in the absence of current saturation is introduced, followed by a discussion of the trajectories under the three types of CSAs. The analysis in this section is based on the model in Fig.~\ref{fig:Single Machine Grid Connected Circuit Model}, which is the equivalent model shown in Fig.~\ref{fig:System_Model}(a).
\vspace{-0.8em}
\vspace{-1em}
\subsection{Current Unsaturation}
\begin{figure}[t]
    \vspace{-0.5em}
    \centering
    \includegraphics[width=0.95\columnwidth]{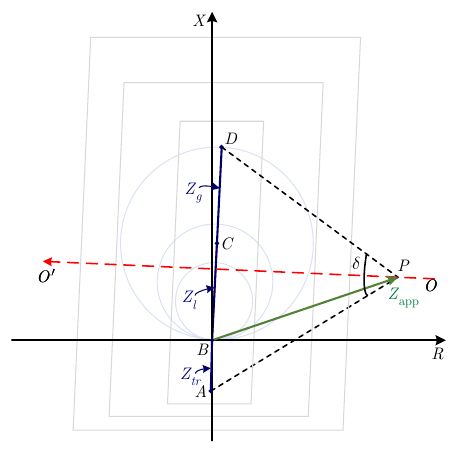}
    \captionsetup{justification=raggedright,singlelinecheck=false} 
    \caption{Power swing trajectory in the absence of current limitation.}
    \label{fig:Unsat_Trajectory}
\end{figure}

Under the assumptions that there is no current limitation in the control loops and that \(|\dot{V}|=|\dot{V}_{g}|\), the apparent impedance observed by the relay near Bus B, marked in red in Fig.~\ref{fig:Single Machine Grid Connected Circuit Model}, is consistent with that in a SG-based system, as derived in~\cite{kundur1994}:
\begin{equation}
\label{eq:Apparent Impedance Unsaturated}
Z_{\text{app}} = \left( Z_{l} + Z_{g} - \frac{1}{2}Z_{T} \right) - j\frac{1}{2}Z_{T}\cot\frac{\delta}{2}.
\end{equation}
Accordingly, with point B as the origin, the trajectory of the apparent impedance on the impedance plane is shown in Fig.~\ref{fig:Unsat_Trajectory}. The green vector \(\overrightarrow{BP}\) represents the apparent impedance measured by the relay, and \(\angle APD\) reflects the power angle \(\delta\) between the GFM IBR and the grid. As \(\delta\) increases from \(0\) to \(2\pi\), the apparent impedance endpoint \(P\) moves along \(OO'\) from right to left. The trajectory \(OO'\) is the perpendicular bisector of the line segment \(AD\), which represents the total impedance between the PCC and the equivalent voltage source of the grid. Therefore, \(|DP|=|AP|\). The figure illustrates that the apparent impedance \(\overrightarrow{BP}\) is uniquely mapped to the power angle \(\delta\), and this relationship can be effectively represented on the impedance plane. The conventional power swing detection scheme is based on this well-established property.
\subsection{Current Saturation with CSAs}
\begin{figure}[t]
    \vspace{-0.5em}
    \centering
    \includegraphics[width=0.98\columnwidth]{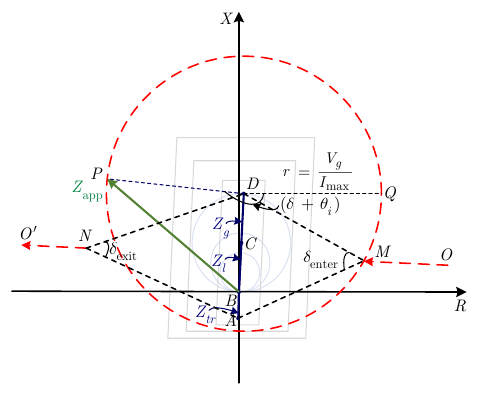}

    \captionsetup{justification=raggedright,singlelinecheck=false} 
    \caption{Power swing trajectory under the CSAs.}
    \label{fig:Saturated_Trajectory} 
\end{figure}

Under the influence of CSAs in the control loops, the apparent impedance in the current saturation mode is
\begin{align}
\label{eq:Apparent Impedance Saturated Variable_1}
Z_{\text{app}} &= \frac{\left( Z_{l} + Z_{g}\right)I\angle \theta_{i} + V_{g}\angle-\delta}{I\angle \theta_{i}}  \\
\label{eq:Apparent Impedance Saturated Variable_2}
&= \left( Z_{l} + Z_{g} \right) + \frac{V_{g}}{I}\angle \left(-\delta-\theta_{i}\right),
\end{align}
where \(\theta_{i}\) represents the phase angle of the saturated current. The apparent impedance trajectory on the impedance plane forms a circle centred at point \(D\) which is the endpoint of the vector \(\left(Z_{l}+Z_{g}\right)\), with a radius \(r=V_{g}/I_{\text{max}}\), i.e., \(DQ\), as shown in Fig~\ref{fig:Saturated_Trajectory}.
When the current does not reach the overcurrent limitation \(I_{\text{max}}\) and remains in the unsaturated mode, the apparent impedance trajectory still follows the line described in the previous subsection, moving along either \(OM\) or \(NO'\). Both \(OM\) and \(NO'\), as well as their extensions, are perpendicular bisectors of the line segment \(AD\). As \(\delta\) gradually increases to \(\delta_{\text{enter}}\), which is \(\angle AMD\), the current reaches the critical saturation state, and the trajectory moves to point \(M\). With further increase in \(\delta\), the current enters the saturation mode, and the trajectory follows the circular path where point \(M\) is located.

The process by which the trajectory transitions from the critical saturation point \(M\) to the saturation trajectory on the circle varies depending on the CSAs.
\subsubsection{Circular CSA}
Once the d-axis and q-axis voltages fail to track their references, the integrators in the voltage PI controllers will clamp soon as introduced in Section III B. Consequently, the magnitudes of \(\bar{i}_{sd}\) and \(\bar{i}_{sq}\) become equal; however, their signs depend on the system parameters, which determine the value of \(\theta_{i}\). As reflected in the impedance plane, the trajectory will quickly move from point \(M\) along the circular path to a specific point on the circle, characterised by a central angle of \(\angle QDP=\delta_{\text{enter}}+\theta_{i}\). Subsequently, it moves clockwise along the circular path as \(\delta\) increases. During saturation, due to the presence of anti-windup, \(\bar{i}_{sd}\) and \(\bar{i}_{sq}\) may change signs, resulting in a change in \(\theta_{i}\) in a short time on the apparent impedance trajectory.
\subsubsection{D-Axis Priority CSA}
After entering the saturation mode, \(\theta_{i}\) undergoes a transient process to reach \(0^\circ\). Subsequently, \(\theta_{i}\) remains at \(0^\circ\) until the GFM IBR exits the saturation mode.
\subsubsection{Q-Axis Priority CSA}
During forced saturation, the q-axis voltage may reach the reference value, resulting in a change in the sign of \(\bar{i}_{sq}\). This results in a \(180^\circ\) shift in \(\theta_{i}\), which is reflected in the impedance trajectory as a change with a central angle of \(180^\circ\) along the circular path.
\section{Optimal Current Saturation Strategy}
\label{Optimal Current Saturation Strategy}
The three typical CSAs may lead to rapid changes in power swing trajectories. If such rapidly changing trajectories pass through the blinders, there is a potential risk of malfunction in power swing detection. To mitigate this risk, an optimal current saturation strategy based on a user-defined constant CSA is proposed in this section.

\begin{figure}[htbp]
    \vspace{-0.5em}
    \centering
    \includegraphics[width=0.95\columnwidth]{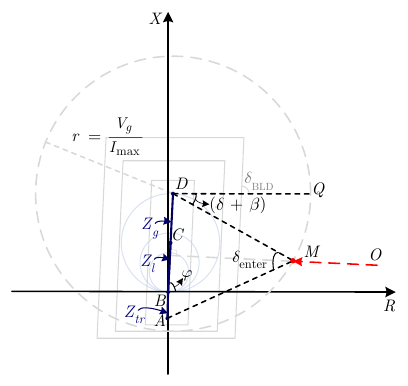}

    \captionsetup{justification=raggedright,singlelinecheck=false} 
    \caption{Condition for the optimal current saturation strategy: the power swing trajectory is continuous at point \(M\).}
    \label{fig:Optimal_Current_Saturation_Strategy} 
\end{figure}

The user-defined constant CSA is used to determine the current saturation angle \(\beta\), represented by \(\theta_{i}\) in~(\ref{eq:Apparent Impedance Saturated Variable_2}). The current saturation angle \(\beta\) governs the starting point of the trajectory once the current enters saturation during a power swing. The optimal strategy is illustrated in Fig~\ref{fig:Optimal_Current_Saturation_Strategy}. 
When the starting point of the saturated trajectory is located at the critical position, i.e., Point M in Fig.~\ref{fig:Optimal_Current_Saturation_Strategy}, the power swing follows a continuous trajectory rather than exhibiting a rapid change. Therefore, Point M is the optimal starting point of the trajectory. The angle \(\angle QDM\) is \((\delta+\beta)\). Then, the optimal saturation angle \(\beta\) for the user-defined constant CSA satisfies the condition that
\begin{equation}
\label{eq:}
(\delta_{\text{enter}}+\beta)+\varphi+(90^{\circ}-\frac{\delta_{\text{enter}}}{2})=180^{\circ},
\end{equation}
where \(\varphi\) is the impedance angle of \((Z_{l}+Z_{g})\). The optimal current saturation angle is
\begin{equation}
\label{eq:optima angle}
\beta_{\text{opt}} = 90^{\circ} - \varphi - \frac{\delta_{\text{enter}}}{2}.
\end{equation}
The second-order rotor dynamics in the GFM-IBR are reflected through the tuning of parameters, which is controllable. Therefore, a continuous power swing trajectory enabled by an appropriate current saturation angle \(\beta\) can help prevent detection malfunctions caused by the rapid trajectory changes inherent in the circular, d-axis priority, and q-axis priority CSAs.
\section{Simulations and Case Studies}
\label{Simulations and Case Studies}
In this section, the test system configuration is introduced, followed by case studies to verify the theoretical analysis and illustrate the impacts of CSAs on power swing detection functions and power system transient stability.
\vspace{-0.5em}
\subsection{System Configuration}
\begin{figure}[!ht]
    \centering
    \includegraphics[width=0.98\columnwidth]{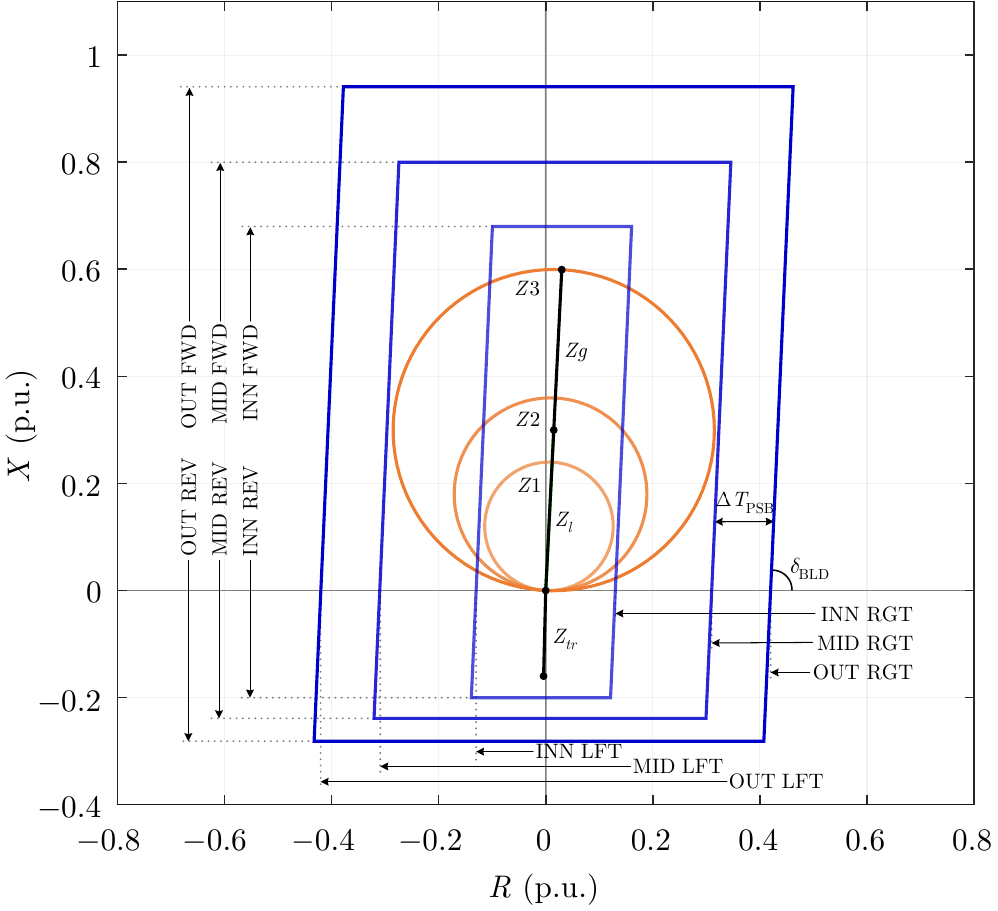}

    \captionsetup{justification=raggedright,singlelinecheck=false} 
    \caption{Three-zone mho characteristics in distance protection and three sets of blinders in power swing detection.}
    \label{fig:Protection_Characteristics} 
\end{figure}

A grid-connected GFM IBR model, as shown in Fig.~\ref{fig:System_Model}, is developed using the MATLAB/Simulink platform to validate the theoretical analysis. The relay, highlighted in red in the figure, employs the conventional three-zone mho characteristics in distance protection, which is shown as the orange circles in Fig.~\ref{fig:Protection_Characteristics}. 
\begin{table}[!ht]
\vspace{0.5em}
\centering
\captionsetup{justification=centering} 
\caption{Parameters of the test system and settings for distance protection and power swing detection}
\begin{tabularx}{\columnwidth}{p{1.3cm} X p{1.9cm} } 
\hline
\multicolumn{1}{c}{\textbf{Parameters}} & \multicolumn{1}{c}{\textbf{Description}} & \multicolumn{1}{c}{\textbf{Value}} \\ \hline
\multicolumn{3}{c}{System Configuration}    \\ 
\hline
\(P_{0}\)          & Active power set point        & 0.6 p.u.  \\
\(f_{n}\)          & Nominal frequency             & 60 Hz  \\
\(v_{d}^{\text{ref}}\)   & PCC voltage set point   & 1 p.u. \\
\(Z_{g}\)          & Impedance of the generator    & 0.3\(\angle87.14 \) p.u.  \\ 
\(Z_{l}\)          & Impedance of the line         & 0.3\(\angle87.14 \) p.u.  \\ 
\(Z_{tr}\)         & Impedance of the transformer  & 0.16\(\angle88.57 \) p.u. \\
\(\angle\phi\)     & Impedance angle of \(Z_{T}\)  & $87.44^{\circ}$           \\
\(I_{\text{max}}\) & Overcurrent limitation        & 1.2 p.u.                  \\
\(u_{\text{max}}\) & Maximum magnitude of the output from the voltage PI controllers        & 0.063 p.u.                  \\
\hline
\multicolumn{3}{c}{Distance Protection Setting} \\ 
\hline
Z1  & 80\% of \(Z_{l}\)   & 0.24\(\angle87.44 \) p.u. \\
TD1 & Zone1 time delay    & 0 s                       \\
Z2  & 120\% of \(Z_{l}\)  & 0.36\(\angle87.44 \) p.u. \\
TD2 & Zone2 time delay    & 0.5 s                     \\
Z3  & 200\% of \(Z_{l}\)  & 0.6\(\angle87.44 \) p.u.  \\ 
TD3 & Zone3 time delay    & 1 s                       \\
\hline
\multicolumn{3}{c}{Power Swing Detection Setting}  \\ 
\hline
OUT RGT    & Outer right blinder    & ~0.42 p.u. \\
OUT LFT    & Outer left blinder     & -0.42 p.u. \\
MID RGT    & Middle right blinder   & ~0.31 p.u. \\
MID LFT    & Middle left blinder    & -0.31 p.u. \\
INN RGT    & Inner right blinder    & ~0.13 p.u. \\ 
INN LFT    & Inner left blinder     & -0.13 p.u. \\
OUT FWD    & Outer forward reach    & ~0.94 p.u. \\ 
OUT REV    & Outer reverse reach    & -0.28 p.u. \\
MID FWD    & Middle forward reach   & ~0.80 p.u. \\ 
MID REV    & Middle reverse reach   & -0.24 p.u. \\
INN FWD    & Inner forward reach    & ~0.68 p.u. \\ 
INN REV    & Inner reverse reach    & -0.20 p.u. \\
\(\delta_{\text{BLD}}\)  & The angles of right and left blinders  & $87.14^{\circ}$  \\ 
\(\Delta T_{\text{PSB}}\)    & Power swing detection time threshold   & 0.033 s \\
\hline
\end{tabularx}
\label{tab:Parameters and setting}
\end{table}
In the figure, the three sets of blinders (inner, middle, and outer), represented in blue, achieve both PSB and OST functionalities. The parameters of the system and the protection setting are shown in TABLE~\ref{tab:Parameters and setting}. 

The power swing detection characteristics are configured to operate across all three distance protection zones, corresponding to 80\%, 120\%, and 200\% of \(Z_l\). The detection process begins when the impedance trajectory intersects the outer blinder, at which point a timer is initiated. Once the trajectory reaches the middle blinder, the timer stops, and the elapsed time interval $\Delta T$ is compared with the reference threshold $\Delta T_{\text{PSB}}$. This threshold should be determined from transient stability analysis to reliably distinguish between the fastest stable power swings and faults. If $\Delta T < \Delta T_{\text{PSB}}$, it indicates a fault. Otherwise, if $\Delta T > \Delta T_{\text{PSB}}$, it indicates a power swing, and the PSB function is subsequently activated to block distance protection and prevent unintended tripping. To ensure that the PSB function covers all distance protection zones, the middle blinders should encompass the largest zone, while the outer blinders should be set based on transient stability analysis to avoid overlapping with maximum load impedance. If a power swing is detected and the trajectory intersects the inner blinder, the OST function is triggered. The system will initiate separation at the pre-determined location. Therefore, the inner blinder should be set to allow only unstable swings to pass through. The blinder angles should also align with the characteristic angle of the mho element.

\subsection{Verification of Full-Cycle Impedance Trajectories}
To verify the full-cycle impedance trajectories derived in Section~\ref{Analysis of Power Swing Trajectories}, four simulation cases (A, B, C, and D) are established, with detailed event descriptions provided in TABLE~\ref{tab:Case A to D}.
\begin{table}[htbp]
\centering
\captionsetup{justification=centering}
\caption{The event settings for Cases A, B, C, and D}
\begin{tabularx}{\columnwidth}{>{\raggedright\arraybackslash}p{0.12\columnwidth}  
                                >{\centering\arraybackslash}X 
                                >{\centering\arraybackslash}X 
                                >{\centering\arraybackslash}X 
                                >{\centering\arraybackslash}X}
\hline
\textbf{} & \textbf{Case A} & \textbf{Case B} & \textbf{Case C} & \textbf{Case D} \\
\hline
CSA & No CSA & Circular & D-Priority & Q-Priority \\
Event & \multicolumn{4}{>{\centering\arraybackslash}p{0.75\columnwidth}}{
    A three-phase-to-ground fault occurs at the terminal of the PCC at \(t = 4\,\text{s}\) and is cleared 150 ms later.
} \\

\hline
\end{tabularx}
\label{tab:Case A to D}
\end{table}

\begin{figure}[h]
    \centering
        \captionsetup[subfigure]{justification=centering}
     \begin{subfigure}[b]{0.24\textwidth}
     \centering
        \includegraphics[width=\textwidth]{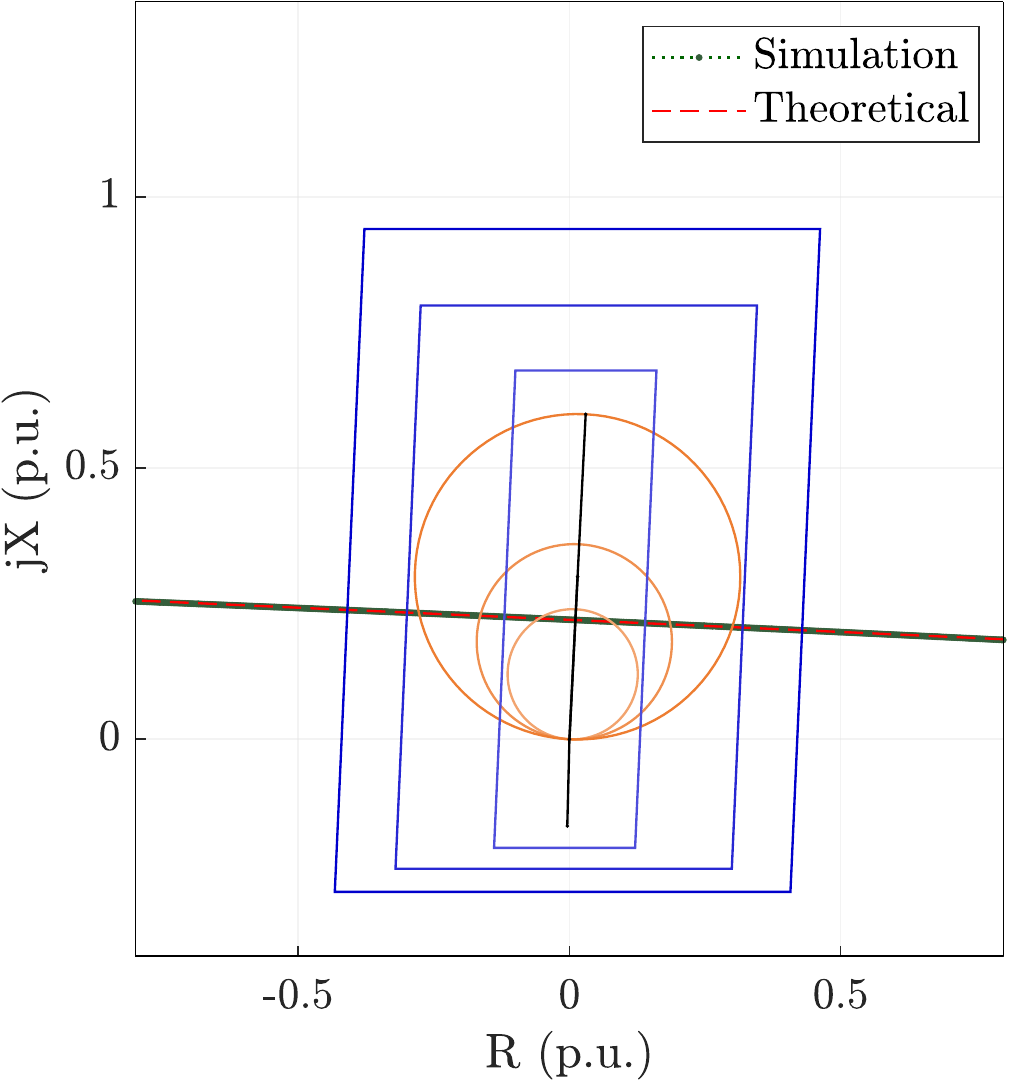}
        \caption{}
        \label{fig:sub1}
    \end{subfigure}
    \hfill
    \begin{subfigure}[b]{0.24\textwidth}
    \centering
        \includegraphics[width=\textwidth]{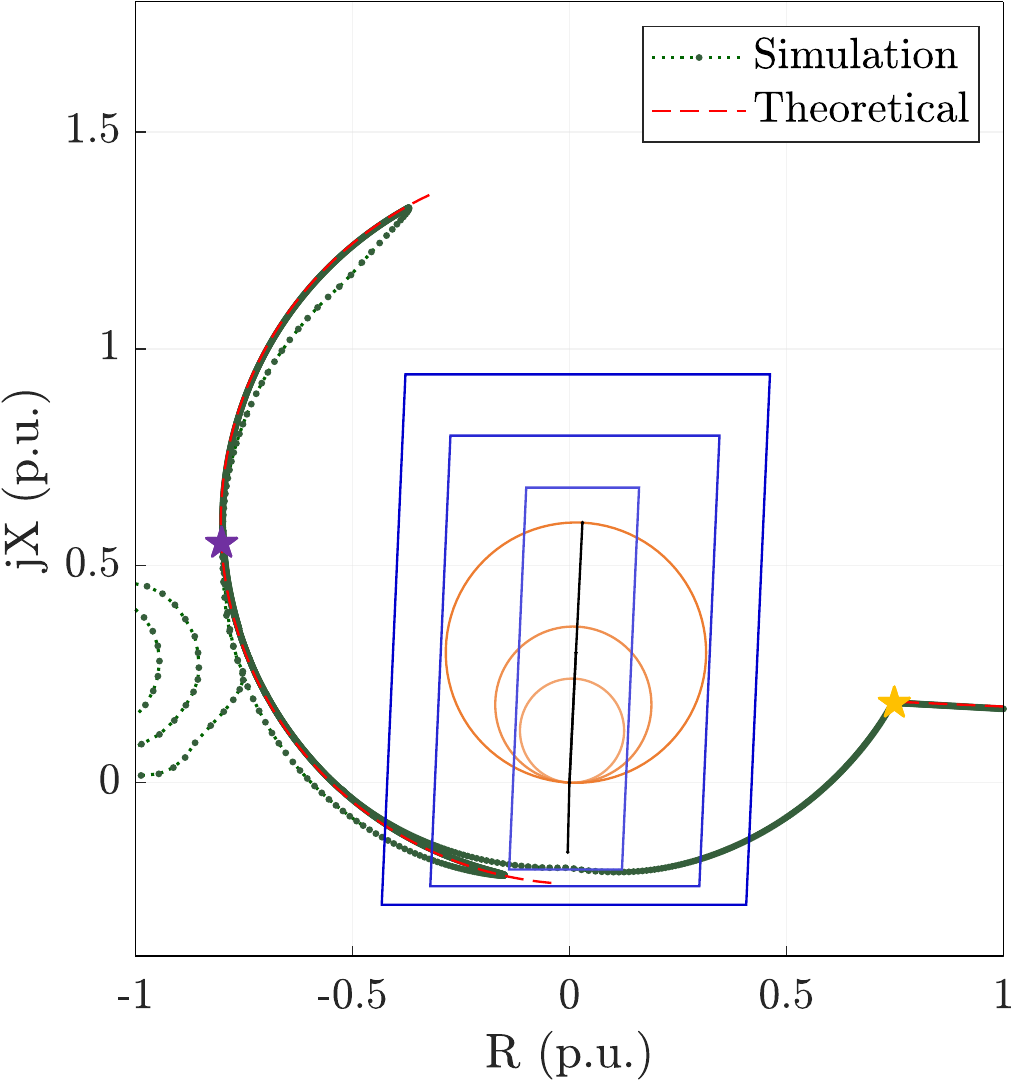}
        \caption{}
        \label{fig:sub2}
    \end{subfigure}
    \hfill
    \begin{subfigure}[b]{0.24\textwidth}
    \centering
        \includegraphics[width=\textwidth]{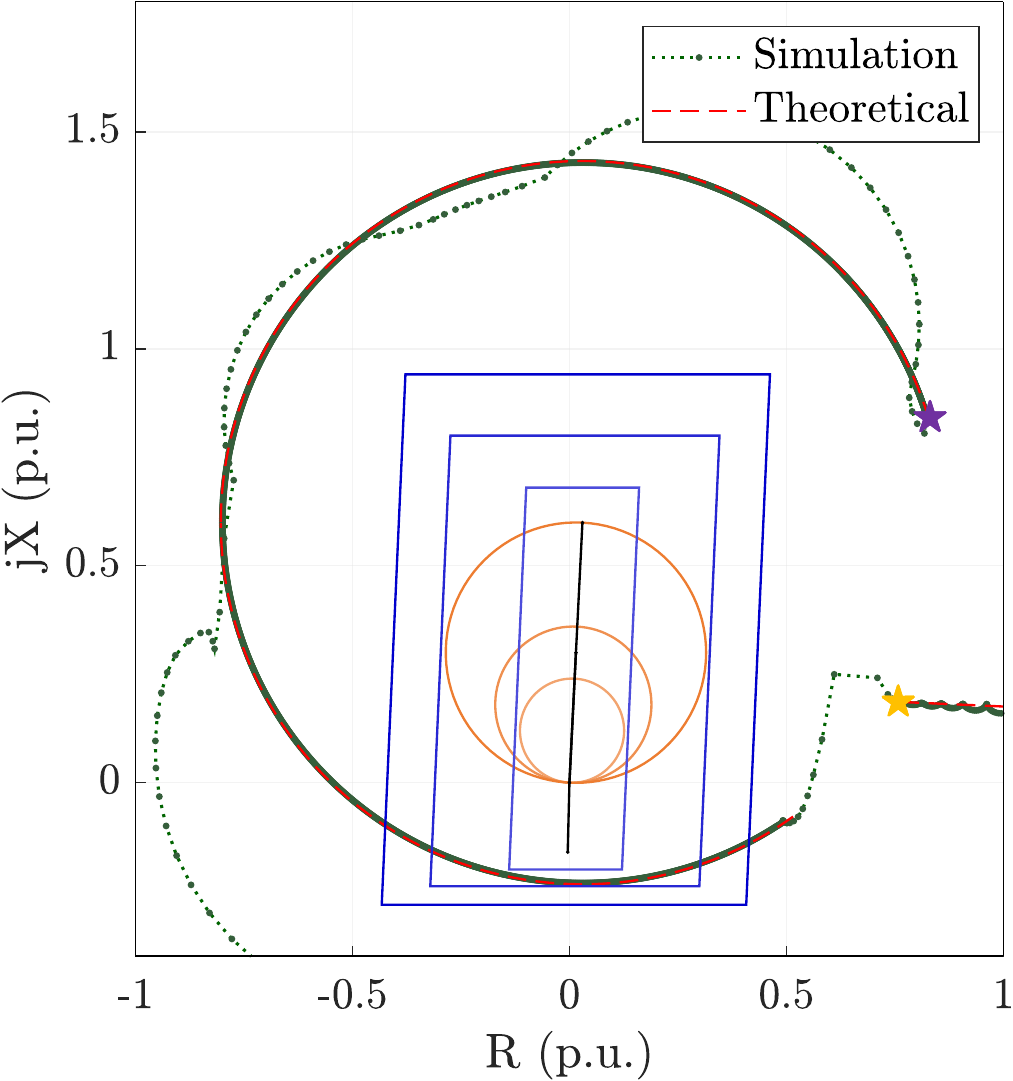}
        \caption{}
        \label{fig:sub3}
    \end{subfigure}
    \hfill
    \begin{subfigure}[b]{0.24\textwidth}
    \centering
        \includegraphics[width=\textwidth]{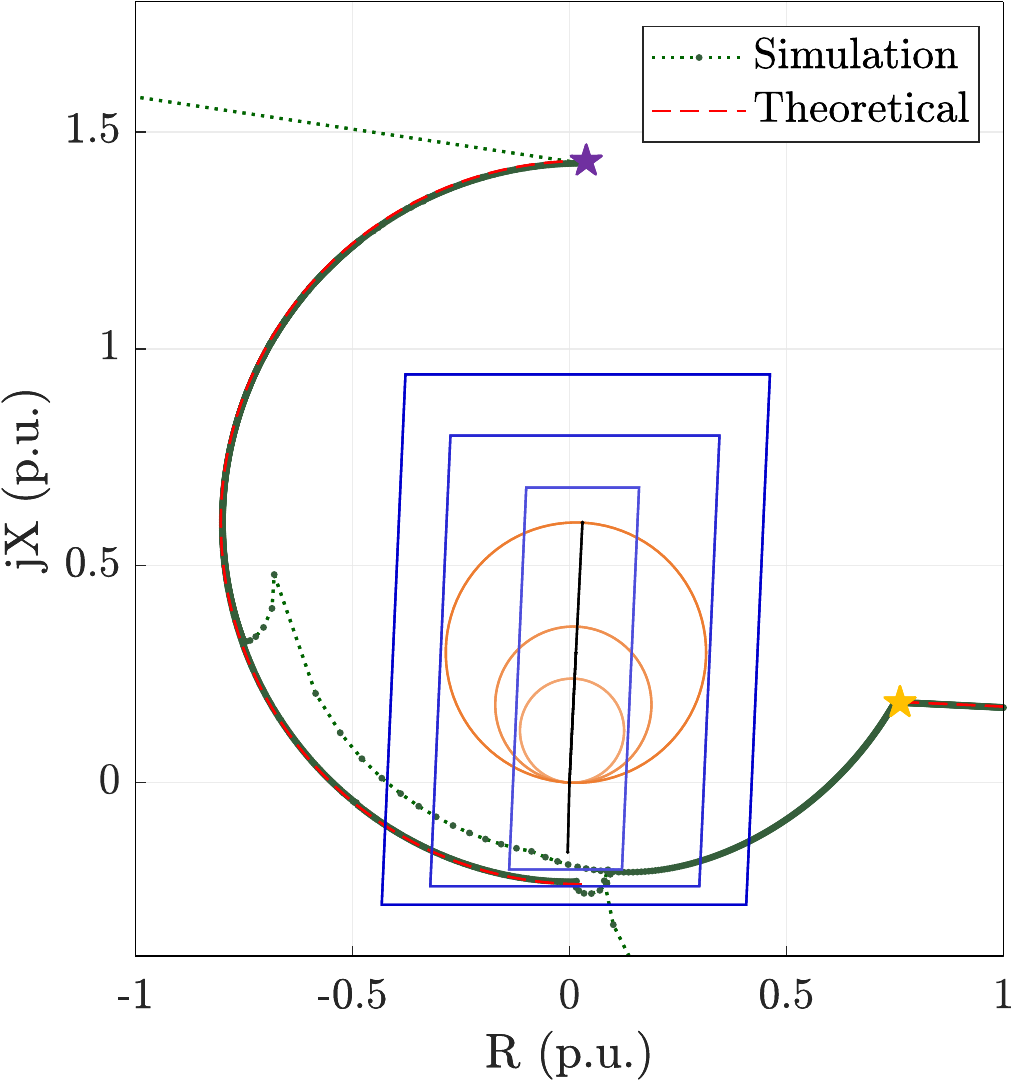}
        \caption{}
        \label{fig:sub4}
    \end{subfigure}
    \caption{The apparent impedance trajectories for Cases A, B, C, and D. The yellow stars indicate the critical points where the current enters saturation, while the purple stars mark where it exits saturation. (a) Case A: Without CSA. (b) Case B: Circular CSA. (c) Case C: D-Axis Priority CSA. (d) Case D: Q-Axis Priority CSA.}
    \label{fig:Case_A_to_D_results}
        \captionsetup{justification=raggedright,singlelinecheck=false}
\end{figure}

The system configuration and other settings are identical to those listed in TABLE~\ref{tab:Parameters and setting}. The relay-observed apparent impedance trajectories throughout an entire swing cycle are illustrated in Fig.~\ref{fig:Case_A_to_D_results}.
The subfigures correspond to the cases without the CSA strategy, with circular, d-axis priority, and q-axis priority CSA, respectively.
After the fault is cleared, all four cases experience unstable power swings. In Fig.~\ref{fig:Case_A_to_D_results}(a), the trajectory of the case without CSA is the same as that of the SG-based system, which is a straight line. The simulation results are consistent with the theoretical analysis. In Fig.~\ref{fig:Case_A_to_D_results}(b), (c), and (d), the trajectories during the current saturation mode are on the circle which are consistent with the theoretical analysis in Section~\ref{Analysis of Power Swing Trajectories}. Cases B, C, and D also demonstrate that under different CSAs, the saturation exiting angles are distinct. These differences are reflected on the impedance plane as variations in the specific points where the trajectories exit the circular path. The locations of these exit points are determined by both \(\delta_{\text{exit}}\) and \(\theta_i\). 
The theoretical saturation entry and exit angles, along with their corresponding simulation results as illustrated in the trajectories of Fig.~\ref{fig:Case_A_to_D_results}, are presented in Table~\ref{tab:Angle calculation}. The yellow stars in Fig.~\ref{fig:Case_A_to_D_results} indicate the points where the GFM-IBR enters the saturation mode, while the purple stars indicate where it exits the saturation mode. The theoretical values show good agreement with the simulation results.
\begin{table}[htbp]
\centering
\captionsetup{justification=centering} 
\caption{Critical angles for entering and exiting saturation under different CSAs in Cases B to D}
\begin{tabularx}{\columnwidth}{>{\raggedright\arraybackslash}X >{\raggedleft\arraybackslash}p{0.2\columnwidth} >{\raggedleft\arraybackslash}p{0.2\columnwidth} >{\raggedleft\arraybackslash}p{0.2\columnwidth}} 
\hline
\textbf{Unit: deg (\(^\circ\))} & 
\multicolumn{1}{r}{\textbf{Circular}} & 
\multicolumn{1}{r}{\textbf{D-Axis Priority}} & 
\multicolumn{1}{r}{\textbf{Q-Axis Priority}} \\
\hline
\multicolumn{4}{c}{Theoretical Values}  \\ 
\hline
\(\delta_{\text{enter}}\)  &  \(54.26\)  &  \(54.26\)  &  \(54.26\)   \\
\(\delta_{\text{exit}}\)   &  \(\pm42.30\)  &  \(\pm16.41\)  &  \(\pm2.34\)   \\
\(\delta_{\text{exit}}+\theta _{i}\)   &  \(182.70\)  &  \(343.59\)  &  \(267.77\)   \\
\hline
\multicolumn{4}{c}{Simulation Results}  \\ 
\hline
\(\delta_{\text{enter}}\)              &  \(54.44\)  &  \(54.66\)  &  \(54.94\)    \\
\(\delta_{\text{exit}}\)               &  \(316.40\)  &  \(342.90\)  &  \(357.50\) \\
\(\delta_{\text{exit}}+\theta _{i}\)   &  \(179.10\)  &  \(343.80\)  &  \(269.80\) \\
\hline
\end{tabularx}
\label{tab:Angle calculation}
\end{table}
Furthermore, the results in Fig.~\ref{fig:Case_A_to_D_results}(b), (c), and (d) verify that the rapid changes in the trajectory are caused by variations in the current phasor angle, which are in turn triggered by sign changes in the \( i_d \) or \( i_q \).
\subsection{The Impact of CSAs on Power Swing Detection}
\begin{figure}[htbp]
    \centering
        \captionsetup[subfigure]{justification=centering}
     \begin{subfigure}[b]{0.242\textwidth}
     \centering
        \includegraphics[width=\textwidth]{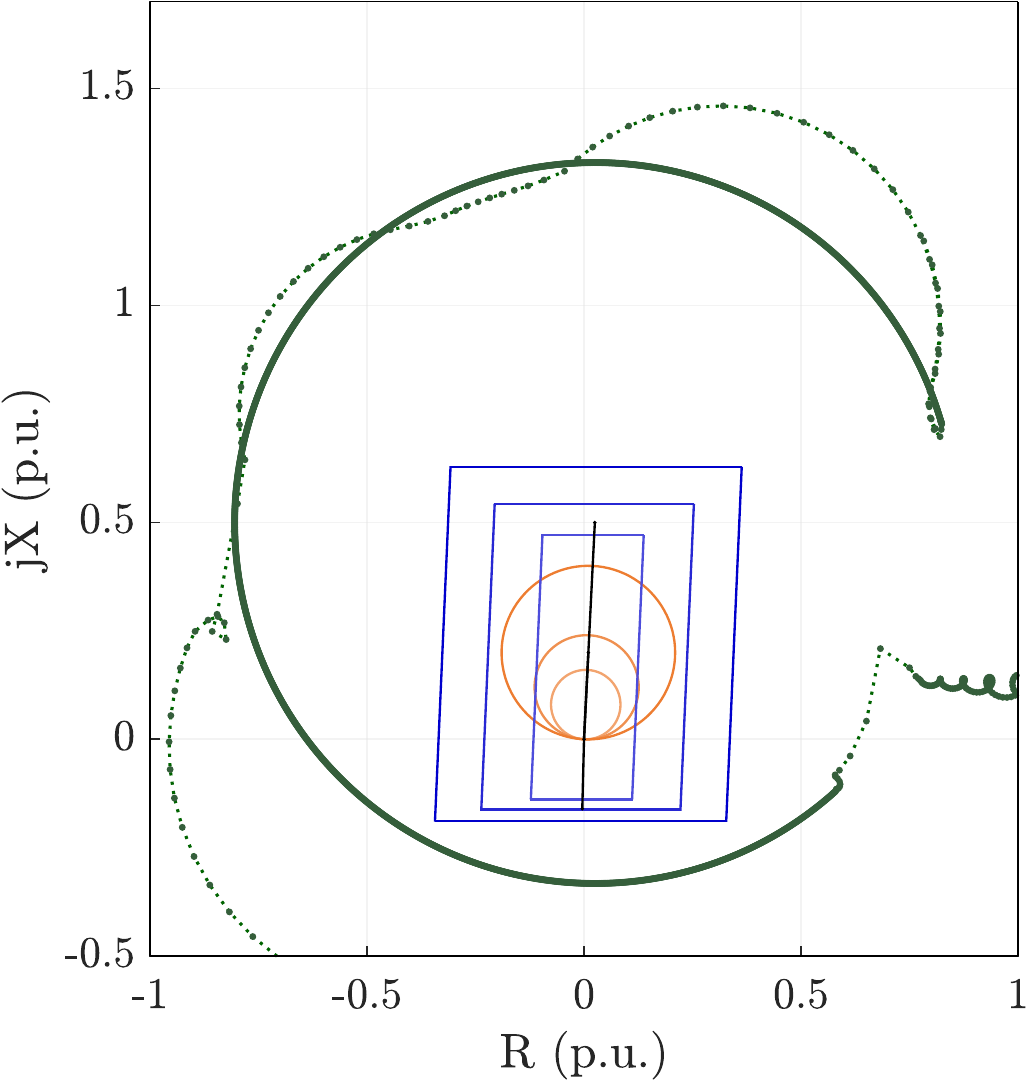}
        \caption{}
        \label{fig:CaseEsub1}
    \end{subfigure}
    \hfill
    \begin{subfigure}[b]{0.24\textwidth}
    \centering
        \includegraphics[width=\textwidth]{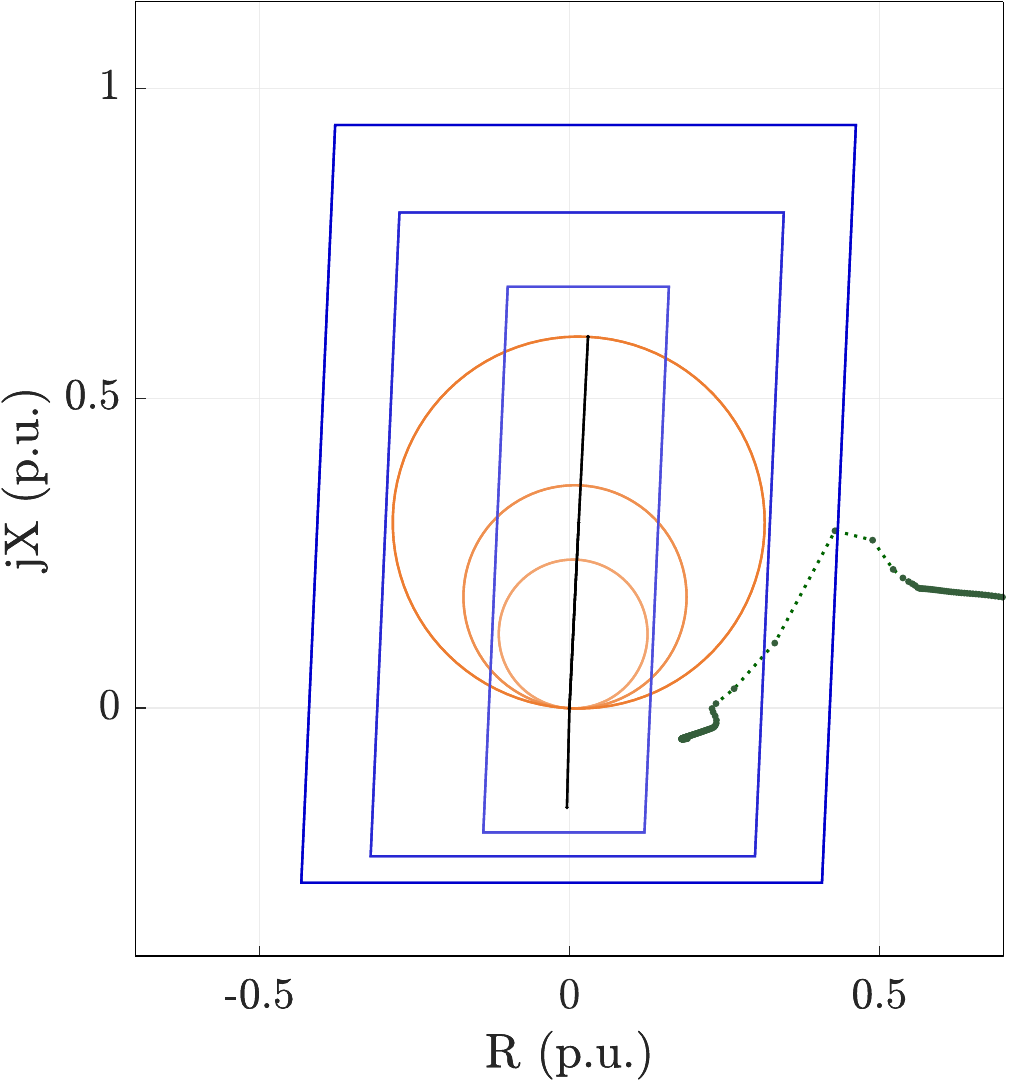}
        \caption{}
        \label{fig:CaseFsub2}
    \end{subfigure}
    \caption{The apparent impedance trajectory of Case E and Case F. (a)Case E. (b)Case F.}
    \label{fig:Case E & F results}
        \captionsetup{justification=raggedright,singlelinecheck=false}
\end{figure}

The new features of apparent impedance trajectories introduced by CSAs can cause power swing detection to malfunction under certain conditions. Cases E and F illustrate two typical scenarios.
\begin{itemize}
\item Case E: D-axis priority CSA. \(Z_{g} = 0.2\,\mathrm{p.u.}\); \(Z_{l} = 0.2\,\mathrm{p.u.}\). A three-phase-to-ground fault occurs at \(t = 4\,\text{s}\) and then is cleared 150~ms later. The power swing detection settings are adjusted following the guidelines outlined in Section~\ref{Simulations and Case Studies}.A.
\item Case F: D-axis priority CSA. \(P_{0} = 0.1\,\mathrm{p.u.}\); \(I_{\text{max}}=1.5\,\mathrm{p.u.}\). Phase jump of \(-78.49 ^\circ\) at \(t = 4\,\text{s}\).
\end{itemize}
The other conditions and settings remain consistent with those in TABLE~\ref{tab:Parameters and setting}. Fig.~\ref{fig:Case E & F results}(a) shows the apparent impedance trajectory of Case E. 
It can be observed that the full-cycle trajectory remains outside the three sets of blinders, meaning that neither stable nor unstable power swings can be correctly detected, resulting in the failure of both PSB and OST. Fig.~\ref{fig:Case E & F results}(b) shows the apparent impedance trajectory in Case F. Under the conditions of Case F, the event is a stable power swing. However, due to the rapid changes in d-axis and q-axis currents caused by the transition process at the early stage of saturation, the time interval for the trajectory passing through the outer and middle blinders is \(\Delta T=0.0056\)~s, which is shorter than \(\Delta T_{\text{PSB}}=0.033\)~s. Consequently, the PSB misidentifies the power swing event as a fault event, leading to a malfunction.
\subsection{Optimal CSA Strategy to Prevent Malfunction}
\begin{figure}[htbp]
    \centering
        \captionsetup[subfigure]{justification=centering}
     \begin{subfigure}[b]{0.24\textwidth}
     \centering
        \includegraphics[width=\textwidth]{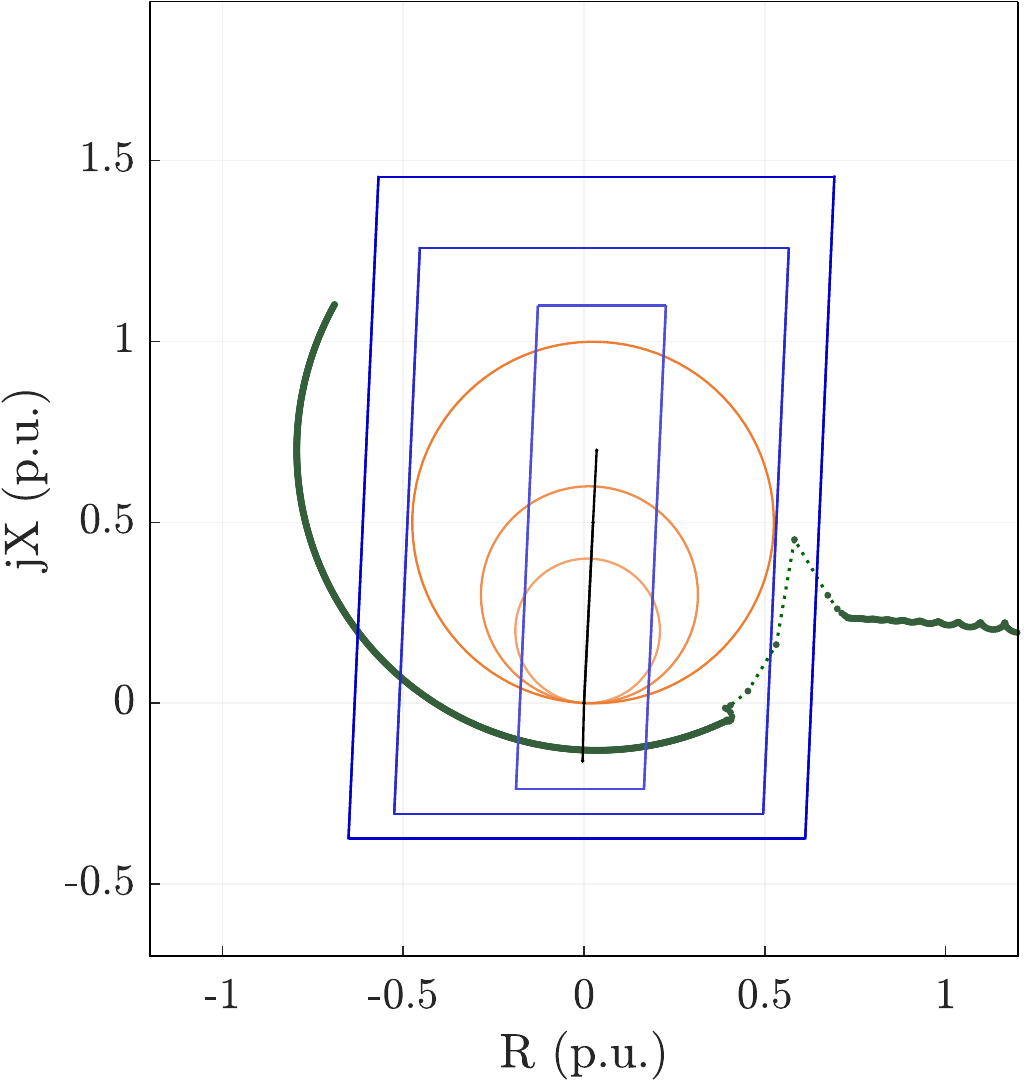}
        \caption{}
        \label{fig:CaseGsub1}
    \end{subfigure}
    \hfill
    \begin{subfigure}[b]{0.24\textwidth}
    \centering
        \includegraphics[width=\textwidth]{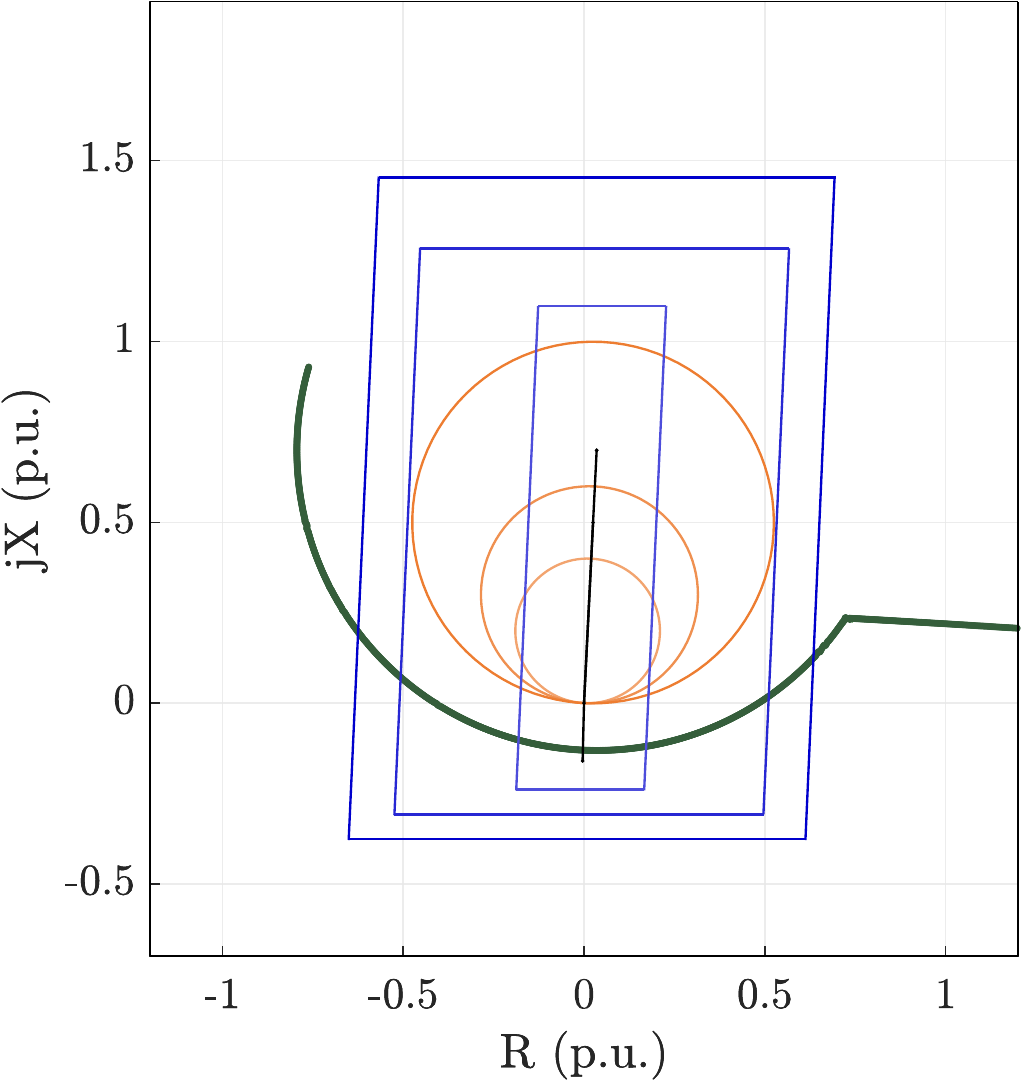}
        \caption{}
        \label{fig:CaseGsub2}
    \end{subfigure}
    \caption{The apparent impedance trajectory of Case G (a)Case G-I. (b)Case G-II.}
    \label{fig:Case G results}
        \captionsetup{justification=raggedright,singlelinecheck=false}
\end{figure}

The optimal CSA, which employs a user-defined constant current saturation angle \(\beta\), is designed to maintain a continuous trajectory during power swing. To verify its effectiveness, Case G-I and Case G-II are presented as follows.
\begin{itemize}
\item Case G-I: D-axis priority CSA. \(Z_{g} = 0.2\,\mathrm{p.u.}\); \(Z_{l} = 0.5\,\mathrm{p.u.}\). \(\Delta P_{0} = +0.5\) at \(t = 4\,\text{s}\). The power swing detection settings are adjusted accordingly.
\item Case G-II: Optimal constant CSA. Under the system condition in this case, the optimal current saturation angle is \(\beta=-32.31^\circ\). The other system configuration and protection settings are the same as Case G-I. 
\end{itemize}
The results of Case G-I and G-II are shown in Fig.~\ref{fig:Case G results}.
Fig.~\ref{fig:Case G results}(a) shows a rapid change in the trajectory when the current enters saturation mode. The time duration during which the trajectory passes through the blinders is \(\Delta t=0.0054\)~s. Fig.~\ref{fig:Case G results}(b), with the optimal constant CSA, exhibits a continuous trajectory. The duration of the trajectory passing through the blinders is 
\(\Delta t=0.043\)~s, which is greater than the setting threshold \(\Delta T_{\text{PSB}}=0.033\)~s for PSB operation.
\subsection{The risk of being unable to exit saturation}
\begin{figure}[t]
\centering
\includegraphics[width=0.98\columnwidth]{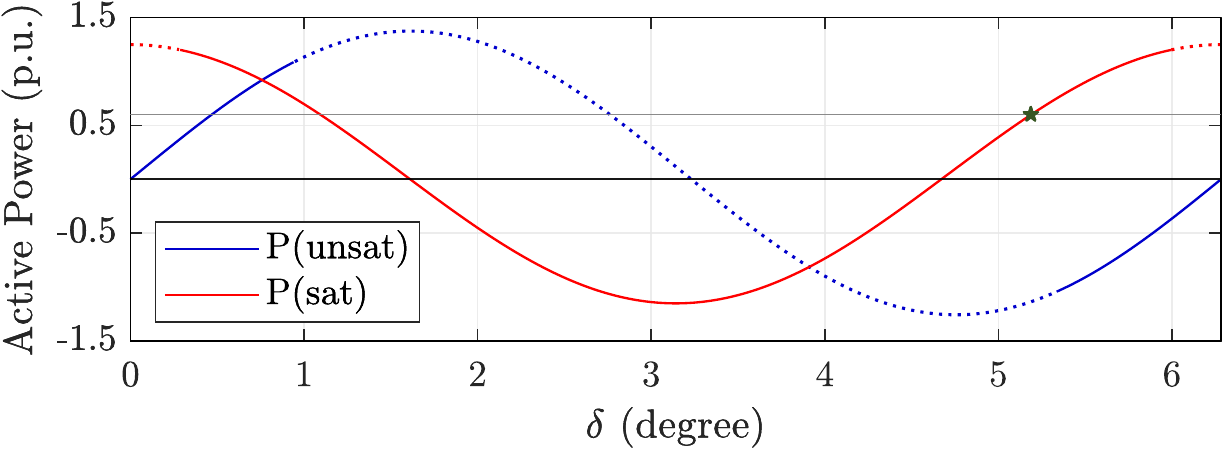}
\caption{$P-\delta$ curves for the d-axis priority CSA. The green star indicates an SSEP.}
\label{fig:P-delta_D-Priority Curve}
\end{figure}

\begin{figure}[htbp]
    \centering
        \captionsetup[subfigure]{justification=centering}
     \begin{subfigure}[b]{0.242\textwidth}
     \centering
        \includegraphics[width=\textwidth]{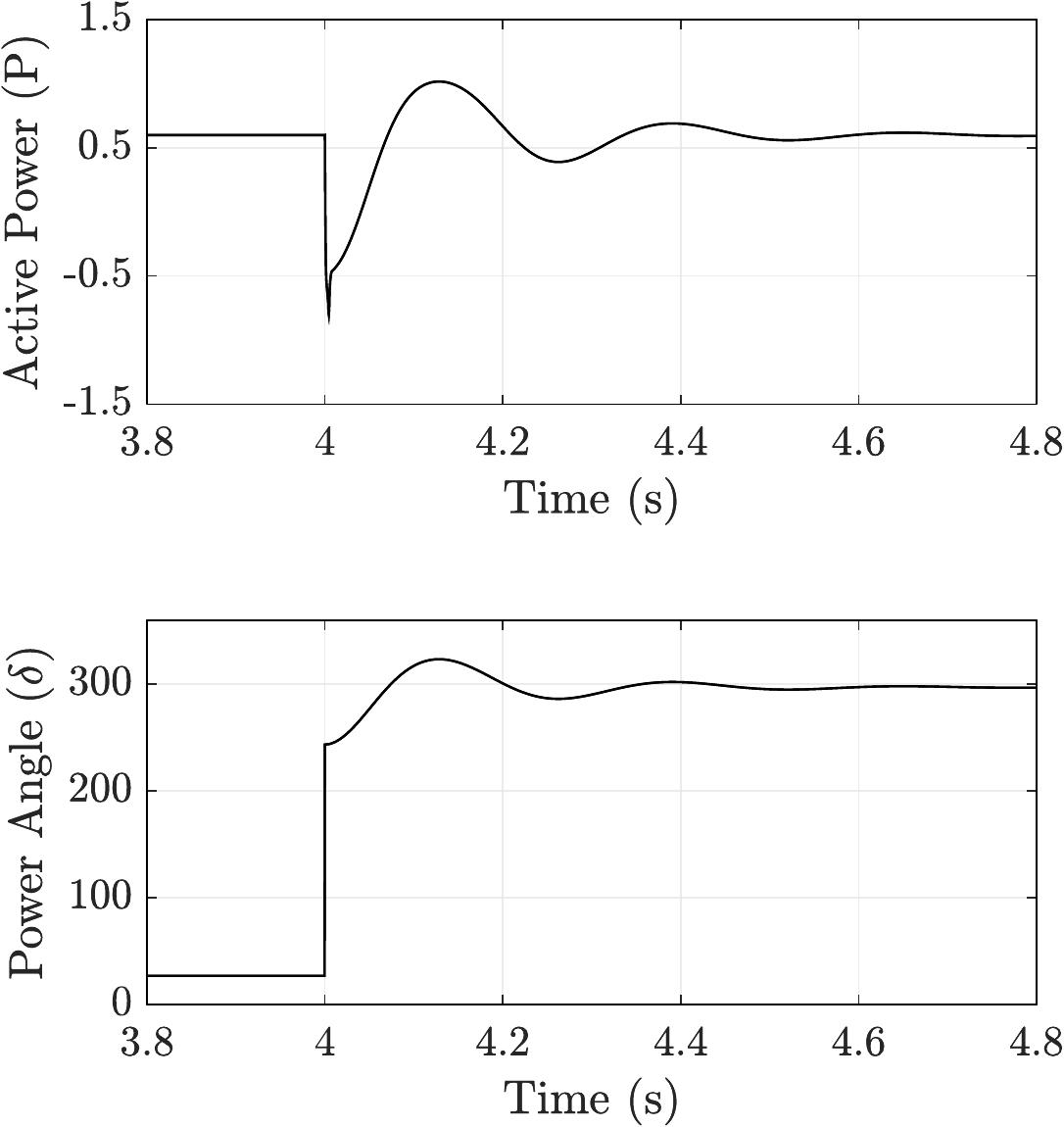}
        \caption{}
        \label{fig:CaseHsub1}
    \end{subfigure}
    \hfill
    \begin{subfigure}[b]{0.235\textwidth}
    \centering
        \includegraphics[width=\textwidth]{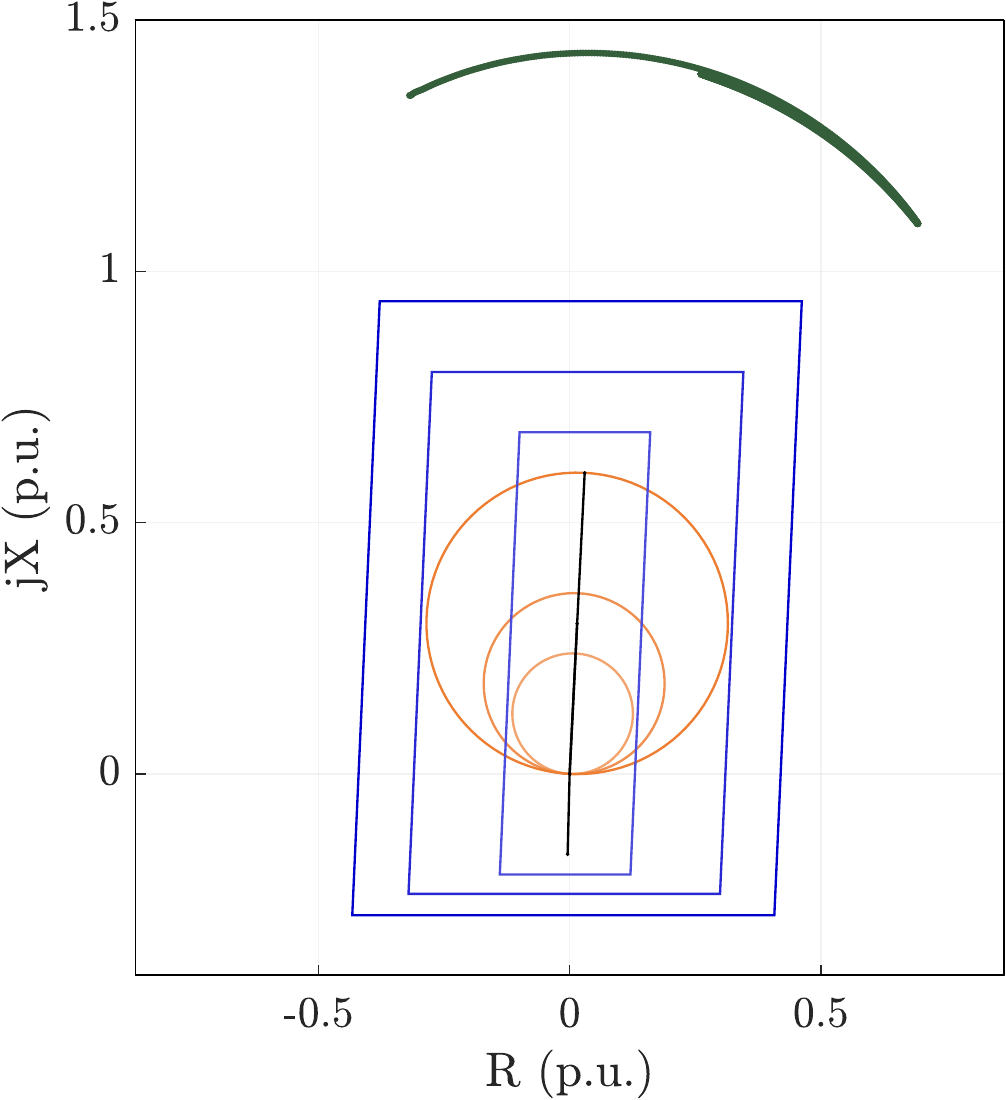}
        \caption{}
        \label{fig:CaseHsub2}
    \end{subfigure}
    \caption{The results of Case H. (a) The curves of active power (p.u.) and power angle (unit: degrees) variations over time. (b) The apparent impedance trajectory.}
    \label{fig:Case H results}
        \captionsetup{justification=raggedright,singlelinecheck=false}
\end{figure}

Different CSAs lead to variations in the \(P-\delta\) curve. For instance, the \(P-\delta\) curve of the d-axis priority CSA is presented in Fig.~\ref{fig:P-delta_D-Priority Curve}. The \(P-\delta\) curve affects not only the transient stability but also the steady-state behaviour of the system. As shown in Fig.~\ref{fig:P-delta_D-Priority Curve}, the green star indicates an SSEP, at which the system remains stable while the current is saturated.
During the fault recovery process, the GFM IBR may become trapped at this SSEP, preventing successful recovery and leading to an undesirable outcome. This scenario is simulated in Case H, with the conditions described as follows.
\begin{itemize}
\item Case H: D-axis priority CSA. Phase jump of \(-216.76^\circ\) at \(t = 4\,\text{s}\).
\end{itemize}
The other conditions remain consistent with those in TABLE~\ref{tab:Parameters and setting}. Fig.~\ref{fig:Case H results} illustrates the results of Case H. 
The active power curve in Fig.~\ref{fig:Case H results}(a) shows that the GFM IBR can output the expected active power. However, the power angle fails to return to the original SEP but instead settles at a new SSEP. The apparent impedance trajectory, shown in Fig.~\ref{fig:Case H results}(b) oscillates along the circular path before eventually stabilising at a point on the circle. This phenomenon is a unique feature introduced by the CSAs of the GFM IBR, which does not exist in conventional power swing detection schemes. 
\section{Conclusion}
\label{Conclusion}
This paper provides a theoretical analysis of the conditions for current entering and exiting saturation under different CSA-based current limiting strategies. Additionally, it investigates how CSAs affect the apparent impedance trajectories, in comparison with those in conventional SG-based systems. The conclusions are summarised as follows.

1) The sets of angles over which circular, d-axis priority, and q-axis priority CSAs enter and exit the current saturation mode are not complementary. They enter saturation under the same angular conditions, but the criteria for exiting saturation differ among the three CSA strategies.

2) Under certain conditions, the circular power swing trajectory introduced by the circular, d-axis priority, and q-axis priority CSAs may cause both PSB and OST to fail, as the trajectory might bypass the blinders during the power swing. Furthermore, the rapid trajectory variations during the transient process of entering current saturation may also lead to malfunction of the PSB.

3) By employing the constant CSA and appropriately setting the current saturation angle \(\beta\), the power swing trajectory can be maintained continuously throughout the transition from the unsaturated to the saturated mode. This continuity effectively mitigates the risk of malfunction in power swing detection caused by rapidly changing trajectories.

4) The d-axis priority CSA may cause the GFM IBR to become trapped in an SSEP. In this scenario, the current cannot exit saturation, leading the apparent impedance trajectory to stabilise at a point on the circular path. This is a unique feature introduced by the CSA-based current limiting strategy.

\footnotesize
\bibliographystyle{IEEEtran}
\bibliography{IEEEabrv,main}

@book{iea2024,
  author    = "{International Energy Agency}",
  title     = "{World Energy Outlook 2024}",
  year      = {2024},
  publisher = "IEA",
  address   = "Paris",
  url       = "https://www.iea.org/reports/world-energy-outlook-2024",
  note      = "Licence: CC BY 4.0 (report); CC BY NC SA 4.0 (Annex A)"
}

@ARTICLE{lasseter2020,
  author={Lasseter, Robert H. and Chen, Zhe and Pattabiraman, Dinesh},
  journal={IEEE Journal of Emerging and Selected Topics in Power Electronics}, 
  title={Grid-Forming Inverters: A Critical Asset for the Power Grid}, 
  year={2020},
  volume={8},
  number={2},
  pages={925-935},
  doi={10.1109/JESTPE.2019.2959271}}

@techreport{lin2020research,
  title={Research roadmap on grid-forming inverters},
  author={Lin, Yashen and Eto, Joseph H and Johnson, Brian B and Flicker, Jack D and Lasseter, Robert H and Villegas Pico, Hugo N and Seo, Gab-Su and Pierre, Brian J and Ellis, Abraham},
  year={2020},
  institution={National Renewable Energy Lab.(NREL), Golden, CO (United States)}
}

@techreport{munz2024,
  title={Protection of 100\% inverter-dominated power systems with grid-forming inverters and protection relays--gap analysis and expert interviews},
  author={Muenz, Ulrich and Bhela, Siddharth and Xue, Nan and Banerjee, Abhishek and Reno, Matthew J and Kelly, Daniel James and Farantatos, Evangelos and Haddadi, Aboutaleb and Ramasubramanian, Deepak and Banaie, Amin},
  year={2024},
  institution={Sandia National Lab.(SNL-NM), Albuquerque, NM (United States)}
}

@inproceedings{Hou2005,
  title={Zero-setting power-swing blocking protection},
  author={Hou, D and Benmouyal, G and Tziouvaras, DA},
  booktitle={3rd IEE international conference on reliability of transmission and distribution networks (RTDN 2005)},
  pages={249--254},
  year={2005},
  organization={IET}
}

@techreport{psrc2005,
  author    = {{IEEE Power System Relaying and Control Committee (PSRC) Working Group WG-D6}},
  title     = {Power swing and out-of-step considerations on transmission lines},
  month     = {Jul.},
  year      = {2005},
  institution = {PSRC},
  url       = {https://bit.ly/3kQx8xI},
  note      = {[Online]}
}

@techreport{nerc2013,
  author    = {{System Protection and Control Subcommittee}},
  title     = {Protection system response to power swings},
  institution = {NERC},
  address   = {Atlanta, USA},
  month     = {Aug.},
  year      = {2013},
  url       = {https://www.bit.ly/3lG62Jv},
  note      = {[Online]}
}

@techreport{haddadi2017,
author = {Haddadi, Aboutaleb and Kocar, Ilhan and Farantatos, Evangelos and Mahseredjian, Jean},
year = {2017},
month = {12},
pages = {},
institution={EPRI},
title = {{EPRI} Technical Report: System Protection Guidelines for Systems with High Levels of Renewables: Impact of Wind \& Solar Generation on Negative-Sequence and Power Swing Protection}
}

@inproceedings{fischer2012tutorial,
  author    = {N. Fischer and G. Benmouyal and D. Hou and D. Tziouvaras and J. Byrne-Finley and B. Smyth},
  title     = {Tutorial on power swing blocking and out-of-step tripping},
  booktitle = {Proceedings of the 39th Annual Western Protective Relay Conference},
  year      = {2012},
  pages     = {1--20}
}

@manual{ge2023,
  author    = {GE},
  title     = {D60 Line Distance Protection System: Instruction Manual},
  address   = {Markham, Canada},
  month     = {Jun.},
  year      = {2023},
  url       = {https://bit.ly/3mucHXR},
  note      = {[Online]}
}

@manual{rel670manual,
  author    = {{Hitachi Energy}},
  title     = {Line Distance Protection REL670, Version 2.2 ANSI: Application Manual},
  series    = {Relion 670 Series},
  year      = {2024},
  url       = {https://bit.ly/3abcXYZ},
  note      = {[Online]}
}

@manual{siemens2023,
  author    = {Siemens},
  title     = {SIPROTEC 5, Distance Protection, 7SA82-V9.5: Manual},
  address   = {Germany},
  month     = {Apr.},
  year      = {2023},
  url       = {https://sie.ag/44MKLPU},
  note      = {[Online]}
}

@article{ieee2016,
  author={IEEE Standards Association},
  journal={IEEE Std C37.113-2015 (Revision of IEEE Std C37.113-1999)}, 
  title={I{EEE} Guide for Protective Relay Applications to Transmission Lines}, 
  year={2016},
  volume={},
  number={},
  pages={1-141},
  doi={10.1109/IEEESTD.2016.7502047}}

@ARTICLE{fan2022,
  author={Fan, Bo and Liu, Teng and Zhao, Fangzhou and Wu, Heng and Wang, Xiongfei},
  journal={IEEE Open Journal of Power Electronics}, 
  title={A Review of Current-Limiting Control of Grid-Forming Inverters Under Symmetrical Disturbances}, 
  year={2022},
  volume={3},
  number={},
  pages={955-969},
  doi={10.1109/OJPEL.2022.3227507}}

@article{qoria2020,
title = {Current limiting algorithms and transient stability analysis of grid-forming VSCs},
journal = {Electric Power Systems Research},
volume = {189},
pages = {106726},
year = {2020},
issn = {0378-7796},
doi = {https://doi.org/10.1016/j.epsr.2020.106726},
url = {https://www.sciencedirect.com/science/article/pii/S0378779620305290},
author = {Taoufik Qoria and François Gruson and Fréderic Colas and Xavier Kestelyn and Xavier Guillaud},
}

@ARTICLE{Rokrok2022,
  author={Rokrok, Ebrahim and Qoria, Taoufik and Bruyere, Antoine and Francois, Bruno and Guillaud, Xavier},
  journal={IEEE Transactions on Power Systems}, 
  title={Transient Stability Assessment and Enhancement of Grid-Forming Converters Embedding Current Reference Saturation as Current Limiting Strategy}, 
  year={2022},
  volume={37},
  number={2},
  pages={1519-1531},
  doi={10.1109/TPWRS.2021.3107959}}

@ARTICLE{fan2023,
  author={Fan, Bo and Wang, Xiongfei},
  journal={IEEE Transactions on Power Systems}, 
  title={Fault Recovery Analysis of Grid-Forming Inverters With Priority-Based Current Limiters}, 
  year={2023},
  volume={38},
  number={6},
  pages={5102-5112},
  doi={10.1109/TPWRS.2022.3221209}}

@ARTICLE{li2023,
  author={Li, Yujun and Lu, Yiyuan and Yang, Jialun and Yuan, Xiaotian and Yang, Rui and Yang, Songhao and Ye, Hua and Du, Zhengchun},
  journal={IEEE Transactions on Energy Conversion}, 
  title={Transient Stability of Power Synchronization Loop Based Grid Forming Converter}, 
  year={2023},
  volume={38},
  number={4},
  pages={2843-2859},
  doi={10.1109/TEC.2023.3283396}}

@article{lu2024,
title = {Segmental equal area criterion for grid forming converter with current saturation},
journal = {International Journal of Electrical Power \& Energy Systems},
volume = {159},
pages = {110015},
year = {2024},
issn = {0142-0615},
doi = {https://doi.org/10.1016/j.ijepes.2024.110015},
url = {https://www.sciencedirect.com/science/article/pii/S0142061524002369},
author = {Yiyuan Lu and Yujun Li and Tongpeng Mu and Chong Shao and Jingrui Liu and Dongmei Yang and Zhengchun Du},
}

@ARTICLE{arjomandi2024,
  author={Arjomandi-Nezhad, Ali and Guo, Yifei and Pal, Bikash C. and Yang, Guangya},
  journal={IEEE Transactions on Energy Conversion}, 
  title={Modeling Fault Recovery and Transient Stability of Grid-Forming Converters Equipped With Current Reference Limitation}, 
  year={2024},
  volume={},
  number={},
  pages={1-13},
  doi={10.1109/TEC.2024.3507544}}

@ARTICLE{Haddadi2019,
  author={Haddadi, Aboutaleb and Kocar, Ilhan and Karaagac, Ulas and Gras, Henry and Farantatos, Evangelos},
  journal={IEEE Transactions on Power Delivery}, 
  title={Impact of Wind Generation on Power Swing Protection}, 
  year={2019},
  volume={34},
  number={3},
  pages={1118-1128},
  doi={10.1109/TPWRD.2019.2896135}}

@article{Haddadi2021,
author = {Haddadi, Aboutaleb and Farantatos, Evangelos and Kocar, Ilhan and Karaagac, Ulas},
year = {2021},
month = {02},
pages = {1050},
title = {Impact of Inverter Based Resources on System Protection},
volume = {14},
journal = {Energies},
doi = {10.3390/en14041050}
}

@ARTICLE{Rao2022,
  author={Rao, J. Tejeswara and Bhalja, Bhavesh R. and Andreev, Mikhail V. and Malik, Om P.},
  journal={IEEE Transactions on Power Delivery}, 
  title={Synchrophasor Assisted Power Swing Detection Scheme for Wind Integrated Transmission Network}, 
  year={2022},
  volume={37},
  number={3},
  pages={1952-1962},
  doi={10.1109/TPWRD.2021.3101846}}

@inproceedings{Jayamohan2023,
  title={Impedance Trajectories during Stable and Unstable Power Swings in Presence of PQ Control based PV Generations},
  author={Jayamohan, Meenu and Das, Sarasij and Brahma, Sukumar},
  booktitle={2023 IEEE Power \& Energy Society General Meeting (PESGM)},
  pages={1--5},
  year={2023},
  organization={IEEE}
}

@ARTICLE{nasr2024part1,
  author={Nasr, Mohamad-Amin and Hooshyar, Ali},
  journal={IEEE Transactions on Power Delivery}, 
  title={Power Swing in Systems With Inverter-Based Resources—Part I: Dynamic Model Development}, 
  year={2024},
  volume={39},
  number={3},
  pages={1889-1902},
  doi={10.1109/TPWRD.2024.3382814}}

@ARTICLE{nasr2024part2,
  author={\vspace{0em}Nasr, Mohamad-Amin and Hooshyar, Ali},
  journal={IEEE Transactions on Power Delivery}, 
  title={Power Swing in Systems With Inverter-Based Resources—Part II: Impact on Protection Systems}, 
  year={2024},
  volume={39},
  number={3},
  pages={1903-1917},
  doi={10.1109/TPWRD.2024.3382843}}

@ARTICLE{xiong2023,
  author={Xiong, Yongxin and Wu, Heng and Li, Yifei and Wang, Xiongfei},
  journal={IEEE Transactions on Power Systems}, 
  title={Comparison of Power Swing Characteristics and Efficacy Analysis of Impedance-based Detections in Synchronous Generators and Grid-following Systems}, 
  year={2024},
  volume={},
  number={},
  pages={1-12},
  doi={10.1109/TPWRS.2024.3469235}}

@inproceedings{xiong2024,
  title={Efficacy Analysis of Power Swing Blocking and Out-of-step Tripping Protection for Grid-Following-VSC Systems},
  author={Xiong, Yongxin and Wu, Heng and Wang, Xiongfei},
  booktitle={2023 8th IEEE Workshop on the Electronic Grid (eGRID)},
  pages={1--5},
  year={2023},
  organization={IEEE}
}

@INPROCEEDINGS{Xiong2023unlimited,
  author={\vspace{0em}Xiong, Y. and Wu, H. and Wang, X.},
  booktitle={22nd Wind and Solar Integration Workshop (WIW 2023)}, 
  title={Efficacy analysis of legacy dual-blinder-based power swing detection scheme in grid-forming VSC-based power system}, 
  year={2023},
  volume={2023},
  number={},
  pages={763-768},
  doi={10.1049/icp.2023.2815}}

@article{wang2023transient,
  title={Transient synchronization stability of grid-forming converter during grid fault considering transient switched operation mode},
  author={Wang, Guangyu and Fu, Lijun and Hu, Qi and Liu, Chenruiyang and Ma, Yanhong},
  journal={IEEE Transactions on Sustainable Energy},
  volume={14},
  number={3},
  pages={1504--1515},
  year={2023},
  publisher={IEEE}
}

@book{kundur1994,
  author    = {P. Kundur},
  title     = {Power System Stability and Control},
  publisher = {McGraw-Hill},
  address   = {New York, NY, USA},
  year      = {1994}
}

@article{yang2024protection,
  title={Protection challenges and solutions for AC systems with renewable energy sources: A review},
  author={Yang, Zhe and Wang, Hongyi and Liao, Wenlong and Bak, Claus Leth and Chen, Zhe},
  journal={Protection and Control of Modern Power Systems},
  volume={10},
  number={1},
  pages={18--39},
  year={2024},
  publisher={PSPC}
}

\clearpage

\ifCLASSOPTIONcaptionsoff
  \newpage
\fi
\end{document}